\documentclass[longauth]{aa}  

\usepackage{graphicx}
\usepackage{txfonts}
\usepackage{amsmath}
\usepackage{amssymb}

\usepackage[separate-uncertainty=true,detect-weight]{siunitx}
\usepackage{microtype}
\usepackage{booktabs}
\usepackage{enumerate}
\usepackage{url}
\usepackage{soul}
\usepackage{graphicx}
\usepackage{txfonts}
\usepackage{xcolor}
\usepackage{subcaption}

\usepackage[colorlinks=true, linkcolor=blue, filecolor=magenta, citecolor=blue, urlcolor=cyan]{hyperref}
\usepackage{cleveref}
\def\msol{\mathrm{M}_{\odot}}
\def\rsol{\mathrm{R}_{\odot}}

\DeclareSIUnit \pc {pc}
\DeclareSIUnit \parsec {parsec}
\DeclareSIUnit \asec {arcsec}
\DeclareSIUnit \pixel {pixel}
\DeclareSIUnit \pixels {pixels}
\DeclareSIUnit \Msun {M_{\odot}}
\DeclareSIUnit \Lsun {L_{\odot}}
\DeclareSIUnit \smass {M_{\star}}
\DeclareSIUnit \dex {dex}
\DeclareSIUnit \mag {mag}
\DeclareSIUnit \pixel {pixel}
\DeclareSIUnit \jansky {Jy}
\DeclareSIUnit \stern {sr}
\DeclareSIUnit \yr {yr}

\begin{document} 

   \title{Rapid formation of a very massive star  $>\!\!50000\,\mathrm{M}_{\odot}$ and subsequently an IMBH from runaway collisions}

   \subtitle{Direct $N$-body and Monte Carlo simulations of dense star clusters}

    \author{
    Marcelo~C.~Vergara \inst{1}\thanks{E-mail: Marcelo.C.Vergara@uni-heidelberg.de (MCV)}  \and 
    Abbas Askar \inst{2}\thanks{E-mail: askar@camk.edu.pl (AA)}  \and 
    Albrecht~W.~H.~Kamlah \inst{3,1} \and 
    Rainer~Spurzem \inst{1,4,5} \and 
    Francesco Flammini Dotti \inst{6,7,1} \and 
    Dominik R.G. Schleicher \inst{8,9} \and 
    Manuel Arca Sedda \inst{10,11,12,13} \and 
    Arkadiusz Hypki \inst{14,2} \and 
    Mirek Giersz \inst{2} \and 
    Jarrod Hurley \inst{15,16} \and 
    Peter~Berczik \inst{2,17} \and 
    Andres Escala \inst{18} \and 
    Nils~Hoyer \inst{3,19,20} \and 
    Nadine~Neumayer \inst{3} \and 
    Xiaoying Pang \inst{21,22} \and \\
    Ataru Tanikawa \inst{23,24} \and 
    Renyue Cen \inst{25,26} \and 
    Thorsten Naab \inst{27} 
    }

    \institute{
    Astronomisches Rechen-Institut, Zentrum f{\"{u}}r Astronomie, M{\"{o}}nchhofstrasse 12-14, D-69120 Heidelberg, Germany \and
    Nicolaus Copernicus Astronomical Center, Polish Academy of Sciences, Bartycka 18, 00-716 Warsaw, Poland \and
    Max-Planck-Institut f{\"{u}}r Astronomie, K{\"{o}}nigstuhl 17, D-69117 Heidelberg, Germany \and
    National Astronomical Observatories, 20A Datun Rd., Chaoyang District, 100101, Beijing, China \and
    Kavli Institute for Astronomy and Astrophysics, Peking University, 5 Yi He Yuan Road, Haidian District, Beijing 100871, China \and
    Department of Physics, New York University Abu Dhabi, PO Box 129188 Abu Dhabi, UAE
\and Center for Astrophysics and Space Science (CASS), New York University Abu Dhabi, PO Box 129188, Abu Dhabi, UAE
\and
    Dipartimento di Fisica, Sapienza Università di Roma, Piazzale Aldo Moro 5, 00185 Rome, Italy 
    \and 
    Departamento de Astronom\'ia, Facultad Ciencias F\'isicas y Matem\'aticas,    Universidad de Concepci\'on, Av. Esteban Iturra s/n Barrio Universitario, Casilla 160-C, Conc
   epci\'on, Chile
   \and
    Gran Sasso Science Institute, Viale F. Crispi 7, I--67100 L'Aquila, Italy \and
    Physics and Astronomy Department Galileo Galilei, University of Padova, Vicolo dell'Osservatorio 3, I--35122, Padova, Italy \and
    INFN – Laboratori Nazionali del Gran Sasso, 67100 L’Aquila, (AQ), Italy \and
    INAF – Osservatorio Astronomico d’Abruzzo, Via M. Maggini snc, 64100 Teramo, Italy \and
    Faculty of Mathematics and Computer Science, A. Mickiewicz University, Uniwersytetu Pozna\'nskiego 4, 61-614 Pozna\'n, Poland \and
    OzGrav: The ARC Centre of Excellence for Gravitational Wave Discovery, Hawthorn, VIC 3122, Australia \and
    Centre for Astrophysics and Supercomputing, Department of Physics and Astronomy, John Street, Hawthorn, Victoria, Australia 3122 \and
    Main Astronomical Observatory, National Academy of Sciences of Ukraine, 27 Akademika Zabolotnoho St, 03143 Kyiv, Ukraine \and
    Departamento de Astronomía, Universidad de Chile, Casilla 36-D, Santiago, Chile \and
    Donostia International Physics Center, Paseo Manuel de Lardizabal 4, E-20118 Donostia-San Sebasti{\'{a}}n, Spain \and
    Universit{\"{a}}t Heidelberg, Seminarstrasse 2, D-69117 Heidelberg, Germany \and
    Department of Physics, Xi'an Jiaotong-Liverpool University, 111 Ren’ai Road, Dushu Lake Science and Education Innovation District, Suzhou 215123, Jiangsu Province, P.R. China \and
    Shanghai Key Laboratory for Astrophysics, Shanghai Normal University, 100 Guilin Road, Shanghai 200234, P.R. China \and
    Department of Earth Science and Astronomy, College of Arts and Sciences, University of Tokyo, 3-8-1 Komaba, Meguro-ku, Tokyo, 153-8902, Japan \and
    Center for Information Science, Fukui Prefectural University, 4-1-1 Matsuoka Kenjojima, Eiheiji-cho, Fukui, 910-1195, Japan \and
    Center for Cosmology and Computational Astrophysics, Institute for Advanced Study in Physics, Zhejiang University, Hangzhou 310027, China \and
    Institute of Astronomy, School of Physics, Zhejiang University, Hangzhou 310027, China \and
    Max Planck Institute for Astrophysics, Karl-Schwarzschild-Str. 1, 85741 Garching, Germany
    }

    \titlerunning{Rapid formation of a VMS and subsequently an IMBH in a dense star cluster}
    \authorrunning{Vergara et al}

   \date{Received September 15, 1996; accepted March 16, 1997}
 
  \abstract
{We present simulations of a massive young star cluster using \textsc{Nbody6++GPU} and \textsc{MOCCA}. The cluster is initially more compact than previously published models, with one million stars, a total mass of $5.86 \times 10^5~\mathrm{M}_{\odot}$, and a half-mass radius of $0.1~\mathrm{pc}$.}
{We analyse the formation and growth of a very massive star (VMS) through successive stellar collisions and investigate the subsequent formation of an intermediate-mass black hole (IMBH) in the core of a dense star cluster.
}
{We use both direct \textit{N}-body and Monte Carlo simulations, incorporating updated stellar evolution prescriptions (SSE/BSE) tailored to massive stars and VMSs. These include revised treatments of stellar radii, rejuvenation, and mass loss during collisions. While the prescriptions represent reasonable extrapolations into the VMS regime, the internal structure and thermal state of VMSs formed through stellar collisions remain uncertain, and future work may require further refinement.}
{We find that runaway stellar collisions in the cluster core produce a VMS exceeding $5 \times 10^4~\mathrm{M}_{\odot}$ within 5 Myr, which subsequently collapses into an IMBH. We stress that further work on stellar astrophysics is needed, particularly in the context of VMS formation, which currently represents important uncertainties.} 
{Our model suggests that dense stellar environments may enable the formation of very massive stars and massive black hole seeds through runaway stellar collisions. These results provide a potential pathway for early black hole growth in star clusters and offer a theoretical context for interpreting recent JWST observations of young, compact clusters at high redshift.}
{}

   \keywords{star clusters: globular clusters--stars: massive--methods: numerical}

   \maketitle
%

\section{Introduction}
\label{sec:introduction}

Most galaxies contain central supermassive black holes (SMBH, $10^6 - 10^9 ~\rm M_\odot $), embedded in a dense nuclear star cluster. Recent observations revealed the existence of even more massive SMBH ($10^9 - 10^{10} ~\rm M_\odot$) during the first billion years after the big bang (see \citealp{Inayoshi2020} and e.g. \citealp{Adamo2024, Harikane2023, Kokorev2023, Kokorev2024, Goulding2023, Ubler2023, Ubler2024, Furtak2024, Matthee2024, Maiolino2024, Juodzbalis2024, Napolitano2025, Akins2025}). This has sparked renewed interest into the question how such SMBHs could grow in such a relatively short time; if they grow via gas or star accretion from lower mass ''seed'' black holes (BHs). Often seed BHs are identified with intermediate mass black holes (IMBH, mass range of $10^2 - 10^5 ~\rm M_\odot$, see for a review \citealp{Greene2020}), but seed BHs may have to be as massive as $10^6 ~\rm M_\odot$ to explain the most massive SMBH in case of (super-)Eddington accretion \citep{Shapiro2005, Schneider2023}.

The formation of central massive BHs from a primordial gas cloud in galactic nuclei has already been postulated by \citet{Rees1984}. Formation channels quoted in his famous Fig.~1 are a dense star cluster undergoing stellar collisions and mergers, a dense cluster of BHs, or a single supermassive star. We summarize the current ideas about these channels here, following \citet{Inayoshi2020} and \citet{Volonteri2021}:\\

	Gravitational mergers between massive stars as a source of IMBH seeds have been discussed in the literature already for decades \citep{Sanders1970,Lee1987,Quinlan1990,PortegiesZwart1999,PortegiesZwart2004, Vanbeveren2009,Giersz2015,Mapelli2016,Reinoso2020,Wang2022, Rantala2024, Fujii2024}. This ``fast'' channel (few Myrs after cluster formation and natal gas expulsion), following \citet{Greene2020}, leads to formation of a very massive star (VMS) by mass segregation and subsequent inelastic collisions of stars in the central region of a dense star cluster. We follow the definition of \citet{Fricke1973,Hara1978,Langbein1990} that a supermassive star (SMS) is characterized by the non-existence of stable nuclear burning, since the Kelvin-Helmholtz contraction leads directly to relativistic collapse; in the first cited paper $5\cdot 10^5\msol$ was given as the limit, but the exact number depends on many parameters; our massive object clearly stays below that limit and is therefore denoted as a VMS. One object outgrows the other ones by growing faster through collisions than its stellar evolution lifetime, which is often called ''runaway'' growth.
    Such condition is not necessary for the collisional fast growth of a massive star. Up to now most works use either relatively small particle number for $N$-body simulations or approximate Fokker-Planck or Monte Carlo models. Until recently, computational limitations made $N$-body simulations with large particle numbers ($\gtrsim 10^6$) unfeasible. The analysis of observational data of nuclear star clusters (NSCs) and the comparison with relevant timescales, further suggested that the collision time in these systems plays a critical role for the formation of their SMBHs \citep{Escala2021}. 
    In order to compare the collision timescale with the age of the system, \citet{Vergara2023} derived a critical mass scale for the formation of massive objects.
    This result is later extended to diverse stellar systems, encompassing various initial conditions, initial mass functions, and evolution scenarios \citep{Vergara2024}. Under the concept of critical mass, \citet{Liempi2025} successfully replicated the observed mass function shape between NSCs and SMBHs with semi-analytic models.

    Gravitational and general relativistic mergers between stellar-mass BHs can lead to the formation of single massive BHs \cite{Zel'dovich1966}, an idea followed up by \cite{Rees1984}. This scenario has been in recent years re-evaluated and re-discovered under the term ``slow'' IMBH growth (around 100 Myr to Gyrs). Mass segregation and stellar evolution generates central BH subsystems, but different from the ideas of \citep{Zel'dovich1966} they are mixed with other stars, and contain binaries, there are not just pure single BH systems. Such models have been shown to generate IMBHs with masses of order $10^2\,\mathrm{M}_{\odot}$ to $10^4\,\mathrm{M}_{\odot}$\citep[e.g.][]{PortegiesZwart2002,Miller2002,Giersz2015,Rodriguez2019,ArcaSedda2019,ArcaSedda2021,ArcaSedda2023,DiCarlo2020,DiCarlo2020a,Kremer2020a,DiCarlo2021,Gonza2021,Leveque2022a,Maliszewski2022,Askar2023,Davis2024,Gonza2024}. Therefore the evolution towards an IMBH depends sensitively on initial parameters such as central density, binary fraction \citep[e.g.][]{Rizzuto2021,Rizzuto2022, ArcaSedda2021,ArcaSedda2021a, Mapelli2022} and there is not necessarily a phase of relativistic collapse of a BH subsystem \citep{Zel'dovich1966}. The presence of gas inflows, for example after galaxy mergers, can substantially contribute to steepen the gravitational potential and enhance the probability for collisions \citep{Davies2011}. As a result, this channel can potentially contribute substantially to explain the SMBH population \citep{Lupi2014, Kroupa2020, Gaete2024}.

    Direct collapse of extremely massive gas clouds can result in massive BHs of order $10^4$~$\mathrm{M}_{\odot}$ to $10^6$~$\mathrm{M}_{\odot}$ effectively bypassing all stellar evolution phases \citep[e.g.][]{Bromm2003,Begelman2006,Begelman2008,Begelman2010, Latif2013, Mayer2015}. However, this channel is difficult to establish, as even trace amounts of dust, ms or formation of molecular hydrogen could cause fragmentation \citep[e.g.][]{Omukai2008, Latif2014, Latif2016}. The resulting objects are expected to either form supermassive stars or quasi-stars (BHs embedded in gaseous envelopes), ultimately collapsing to form massive BHs \citep{Begelman2008, Hosokawa2013, Schleicher2013, Haemmerle2020, Kiyuna2024}. We include here also the possibility of stars that form so massive ($10^5-10^6$~$\mathrm{M}_{\odot}$) that they explode by general-relativistic instability Supernovae (SNe) \citep[e.g.][]{Shibata2002,Sakurai2015,Uchida2017, Ugolini2025}. Under more realistic direct collapse conditions, one may not expect a simple direct collapse, but has to consider the possibility of fragmentation and subsequent mergers of the clumps \citep[e.g.][]{Boekholt2018, Tagawa2020, Schleicher2022, Reinoso2023, Rantala2024, Kritos2024}. The presence of strong far-ultraviolet radiation can allow the formation of stars more massive than $10^4$~$\mathrm{M}_{\odot}$ at $Z \lesssim 10^{-3} Z_\odot$, while for $Z \simeq 10^{-2} Z_\odot$ occurs fragmentation forming dense star cluster that can host stars with $10^3$~$\mathrm{M}_{\odot}$ \citep{Chon2025}.

    Massive Population III (Pop-III) stars have been postulated to produce seed IMBHs with masses of order $10^2$~$\mathrm{M}_{\odot}$ through direct collapse above the pair instability mass gap \citep[e.g.][]{Bromm2004,Bromm2013,Woosley2017,Haemmerle2020, Mestichelli2024}. (Extremely massive) Pop-III stars can merge with other Pop-III stars in their host clusters before collapse to produce even more massive IMBHs above the pair-instability mass gap during the fast gravitational runaway merger phase, as outlined above \citep[e.g.][]{Katz2015,Sakurai2017,Reinoso2018,Reinoso2020,Tanikawa2022,Wang2022, Mestichelli2024}. Collisions between Pop-III stars, which are also accreting material at rates of  $10^{-3}$~$\mathrm{M}_{\odot}\,\mathrm{yr}^{-1}$, can lead to the formation of BHs with masses of the order of $10^4$~$\mathrm{M}_{\odot}$ \citep{Reinoso2025}.\\

Previous work has shown that rapid accretion can lead to very large radii of the supermassive stars forming in such scenarios (up to $100$~AU for $10^4$~M$_\odot$ stars), particularly when the timescale for mass gain is shorter than the Kelvin-Helmholtz timescale of the stars \citep{Hosokawa2013, Schleicher2013, Haemmerle2018, Nakauchi2020}. The formation of VMS has been investigated by \citet{Nakauchi2020} with the MESA code \citep{Paxton2011, Paxton2013, Paxton2015, Paxton2018, Paxton2019} to simulate the evolution of massive stars up to $3000~\rm M_\odot$, accounting for stellar wind mass loss from m-poor to solar metallicity stars. They found that such stars grow from about $30~\rm R_\odot$ at the MS to around $10^4~\rm R_\odot$ in the post-MS phase. Besides \citet{Martinet2023} used the GENEC code \citep{Eggenberger2008}  to model rotating and non-rotating stars of $180$ to $300~\rm M_\odot$ with a metallicity of $Z = 0.014$, finding MS lifetimes of about $2.2-2.6$\,Myr. \\

Globular cluster formation takes place both in the local universe as well as at high redshifts \citep[][see discussion below]{Adamo2024, Mowla2024}. High metallicity clusters may form from high density gas clouds, e.g. in merging galaxies or fragmented tidal arms, and are frequently observed in interacting galaxies. Formation of low metallicity clusters in the early universe, in a cosmological context, is quite difficult regarding the baryonic content of clusters - very narrow metallicity range, which excludes a strong self-enrichment during the formation phase.
Recently, \citet{Kimm2016} succeeded, for the first time, in making m-poor globular clusters at high redshift ($z>7$)
during the epoch of reionization in dwarf galaxies in cosmological simulations. 
These m-poor clusters will later be accreted into  Milky Way - type galaxies. In \citet{Kimm2016}, two out of two simulated dwarf galaxies formed a globular cluster, which is more than sufficient to explain the observed population in the Milky Way. It is even conceivable that still denser and/or more massive globular clusters may be formed through this channel considering that they were modeling two randomly chosen halos. \\

\citet{Adamo2024} found a few compact clusters with intrinsic masses around $10^6\rm~ ~\rm M_\odot$ formed $460$\,Myr after the Big Bang with the \textit{James Webb} Space Telescope (\href{https://webb.nasa.gov}{JWST}). The presence of extremely dense and massive clusters at high redshifts could be essential to explain the high nitrogen-oxygen abundances observed in galaxies \citep{Bouwens2010, Tacchella2023, Marques-Chaves2024, Nagele2023}. VMSs formed in these dense clusters could contribute to these observations through hydrogen-burning nucleosynthesis \citep{Charbonnel2023}. If the VMS is m-enriched, it can emit powerful winds \citep{Sander2020, Vink2022} that can enrich the gas with nitrogen at the expense of oxygen \citep{Glebbeek2009}.
Super-compact massive star clusters can form during galaxy mergers \citep{Renaud2015}, which trigger starburst environments and create clusters with tens of thousands to millions of stars. High-redshift, low-metallicity galaxies \citep{Pettini1997} undergoing mergers are ideal environments for forming massive, compact star clusters where VMS can develop. In conclusion, we are motivated from theory, simulations and observations alike to study the creation of VMSs in dense and massive star clusters that have formed in the early universe. We note that in the local Universe, the most massive stars detected have masses of approximately $100-300~\rm M_\odot$ \citep{Doran2013, Keszthelyi2025}. Beyond the local Universe, stars are too faint to measure their masses directly with current technology.

This paper provides $N$-body models for clusters very similar to the recently observed young massive clusters in the early universe \citep{Mowla2024,Adamo2024}, and shows how they would form a VMS and an IMBH. It gives also sound support by a full and direct star cluster $N$-body simulation for a long-standing conjecture, that VMSs of more than $10^4 \mathrm{M}_{\odot}$ do form quickly and early therein by fast (possibly runaway) merging of normal stars with a massive star, as has been predicted by Monte Carlo simulations (cf. Sect.~\ref{sec:mocca}). Our $N$-body simulations are using \textsc{Nbody6++GPU}, one if not the most advanced and accurate integration code with regard to legacy, astrophysics, numerical and physical accuracy and performance \citep{Aarseth1999,Kamlah2022a,Spurzem2023}. This code is able to routinely simulate a globular cluster size star cluster with all necessary astrophysical ingredients \citep{Wang2015,Wang2016,Kamlah2022a, ArcaSedda2023}.  
While in this work we leverage our code to simulate an unprecedentedly dense and massive star cluster, the huge computational costs limits the number of realisations to one. 
The formation and growth of IMBH by mergers with other BHs will be a source of gravitational wave emission \citep{Rizzuto2021,ArcaSedda2021c,Rizzuto2022}
and tidal disruption events \citep{Stone2016, Rizzuto2023}. 

The paper is organised as follows: our methodology is summarized in Section~\ref{sec:Methods}. The initial conditions for the star clusters are presented in Section~\ref{sec:initialcond}. We analyze our results in Section~\ref{sec:Results}. A summary and conclusions are then presented in Section~\ref{sec:summary}.


\section{Methods}\label{sec:Methods}

In this section, we present the methodology employed, particularly the code \textsc{Nbody6++GPU} in Section~\ref{sec:Nbody6++GPU}, while the the \textsc{MOCCA} code is explained in Section~\ref{sec:mocca}. 

	\subsection{NBODY6++GPU}
  \label{sec:Nbody6++GPU}
	The dense star cluster models are evolved using the state-of-the-art direct force integration code \textsc{Nbody6++GPU}, which is optimised for parallelisation using simultaneously three different levels, bottom-end of many core GPU accelerated parallel computing \citep{Nitadori2012,Wang2015}, mid-level OpenMP thread based parallelisation, and upper level MPI parallelisation \citep{Spurzem1999}. It yields excellent sustained performance on current hybrid massively parallel supercomputers
 \cite[cf.][and Sect.~\ref{sec:Hardware settings}]{Spurzem2023}. It is a successor to the many direct force integration $N$-body codes of gravitational $N$-body problems, which were originally written by Sverre Aarseth \citep[][and sources therein]{Aarseth1985,Aarseth1999,Aarseth1999a,Aarseth2003,Aarseth2008}. The recent review by \citet{Spurzem2023} provides a comprehensive overview over the entire research field of collisional dynamics and how \textsc{Nbody6++GPU} fits within this field. That review also provides some key informations about codes, such as \textsc{PeTaR} \citep{Wang2020a} and \textsc{BiFROST} \citep{Rantala2023}.
\\
The full scale parallel code still retains previously implemented features that are essential for efficient and accurate $N$-body integration on all scales. This includes the Hermite scheme with hierarchical block time-steps \citep{McMillan1986,Hut1995,Makino1991,Makino1992}.
The Ahmad-Cohen (AC) neighbour scheme \citep{Ahmad1973} permits for every star to divide the gravitational forces acting on it into the regular component, originating from distant stars, and an irregular part, originating from nearby stars (''neighbours''). While the AC scheme is very efficient to reduce the number of floating operations necessary, it somewhat disturbs parallelization and acceleration, the code requires both CPU and GPU \citep{Wang2015}. The code also retains Kustaanheimo-Stiefel (KS) regularisation \citep{Stiefel1965}, which allows fast and accurate integration of hard binaries and few-body systems \citep{Mikkola1999,Mikkola1999a,MikkolaAarseth1998} as well as close encounters.
\\  
Post-Newtonian dynamics of relativistic binaries or fast and close hyperbolic encounters is currently still using an orbit-averaged algorithm, for bound systems \citep{Peters1963,Peters1964}; it has been used by different simulations already \citep{DiCarlo2019,DiCarlo2020,DiCarlo2020a,DiCarlo2021,Rizzuto2021,Rizzuto2022,ArcaSedda2021a}, recently also including spin-dependent kicks at mergers \citep{ArcaSedda2023,ArcaSedda2024,ArcaSedda2024a}. For 
unbound systems \citet{Turner1977} \citep[see also more recent][]{Bae2017} the relativistic treatment of hyperbolic encounters is still experimental and not yet public in \textsc{Nbody6++GPU}. However, the use of fully resolved relativistic Post-Newtonian (PN) dynamics inside \textsc{Nbody6} and \textsc{Nbody6++} has been already pioneered by \citet{Kupi2006}, and extended by \citet{Aarseth2012,Brem2013}. A recent paper \cite{Sobolenko2021} shows the current best implementation using PN terms up to order PN3.5 and including spin-spin and spin-orbit interactions. These features also are not yet incorporated in our public code version.

Recently, the 19 \textsc{Dragon-II} simulations, which are formally the successor simulations to the \textsc{Dragon-I} simulations by \citet{Wang2016}, executed with \textsc{Nbody6++GPU} have provided pioneering insights into the evolution and dynamics of young massive star clusters found in the Local Universe as well as binary evolution, IMBH growth and gravitational wave sources within them and in the surrounding field \citep{ArcaSedda2023,ArcaSedda2024,ArcaSedda2024a}. One key update in these simulations next to the stellar evolution published in \citet{Kamlah2022}, has been the inclusion of general relativistic merger recoil kicks \citep{ArcaSedda2023a}. We are using this code version with important updates about stellar collisions, explained in Appendix~\ref{sec:rejuvenation_collisions}. 

\subsection{MOCCA}
\label{sec:mocca}

The fast formation of a VMS through runway collisions of MS stars in dense star clusters has also been demonstrated using star cluster simulation codes based on \citet{Henon1971} Monte Carlo method \citep{Henon1972a,Henon1972b,Henon1975,Stodolkiewicz1982,Stodolkiewicz1986,Giersz1998,Joshi2000,Giersz2001,Giersz2006,Giersz2008,Hypki2013,Rodriguez2022,Giersz2025} for treating stellar dynamics. \citet{Gurkan2004} found that Monte Carlo simulations of star clusters with higher central concentration and short initial relaxation times can undergo core collapse on timescales shorter than the MS lifetime of the most massive stars in the cluster. They found that this early core collapse can lead to an increased collision rate which is likely to result in runaway collisions between MS stars leading to the formation of a VMS. \citet{Freitag2006a,Freitag2006} carried out Monte Carlo simulations of about a 100 star cluster models with varying initial size ($\rm r_{h}$  ranging between $0.03-5 \mathrm{pc}$) and mass ($\rm N_{*}=10^{5}-10^{8}$). These simulations included prescriptions to allow direct collisions between stars, considering both a sticky-sphere assumption for mergers with no mass loss and fractional mass loss of up to 10\%. They found that in their densest cluster models that had core collapse times less than about 3 Myr, runaway mergers between MS stars always led to the formation of a VMS that had mass ranging between $\rm 400 - 4000 \ M_{\odot}$. The studies among others have posited that the VMS that forms in this way could be a potentially  evolve into an IMBH \citep{Freitag2007,Freitag2008,Goswami2012,Gonza2024}.

Simulations of the evolution star cluster models carried out using the Monte Carlo \textsc{MOCCA} code \citep{Hypki2013,Giersz2013} have shown that massive clusters with high central densities ($\rm \rho_{c} \gtrsim 10^{7} \ \mathrm{M}_{\odot} \ pc^{-3}$) are  likely to form a VMS through the runaway collisions of MS stars \citep{Giersz2015,Askar2017,Hong2020,Maliszewski2022,Askar2023}. In order to trigger this pathway for VMS formation and subsequently IMBH formation, mass segregation of the most massive stars in the cluster must occur on a timescale shorter than their lifetime which is about 3 Myr \citep{Gurkan2004}. High central densities and large sizes of MS stars increases their likelihood to collide and form a VMS. From thousands of star cluster models simulated as part of the MOCCA-Survey Database I \citep{Askar2017a}, it has been shown that a rapid collisional runaway leading to IMBH formation always occurs when the initial central density exceeds $\rm \rho_{c} \gtrsim 10^{7} \ \mathrm{M}_{\odot} \ pc^{-3}$ \citep[see Fig. 1 in][]{Hong2020}. The mass of the VMS that forms through this pathway is of the order $10^{3}-10^{4} \ \mathrm{M}_{\odot}$ and depends on the total cluster mass and initial central density. Although the formation of a VMS and subsequently an IMBH through a rapid collisional runaway has been extensively explored in previous \textsc{MOCCA} simulations of massive and dense star clusters, these results had not been directly verified with \textit{N}-body simulations for clusters with extremely high central densities until now.

In addition to employing the \citet{Henon1971} Monte Carlo method to treat two-body relaxation, \textsc{MOCCA} utilizes the \textsc{fewbody} code, a direct \textit{N}-body integrator for small-N gravitational dynamics, to determine the outcomes of close binary-single and binary-binary interactions \citep{Fregeau2004}.
For physical collisions between two MS stars in \textsc{MOCCA}, a sticky-sphere assumption with no mass loss is applied. Similar to \textsc{NBODY6++GPU}, the \textsc{MOCCA} code models stellar and binary evolution using prescriptions from the SSE/BSE rapid population synthesis codes \citep{Hurley2000,Hurley2002}, which have been subsequently improved and extended with more recent developments \citep[see][and references therein]{Kamlah2022a}.
While the \textsc{MOCCA} code can model a galactic tidal field using a point-mass approximation, no external tidal field was applied in the simulations carried out in this study, in order to remain consistent with the direct \textit{N}-body run. A detailed description of the latest \textsc{MOCCA} features is provided in \citet{Hypki2022}. For this work, \textsc{MOCCA} was updated to incorporate the improved treatment of stellar radii and the rejuvenation process for a VMS that have been described in Appendices~\ref{sec:evolution_parameters}, \ref{sec:radius_evolution} and \ref{sec:rejuvenation_collisions}. 
Section~\ref{ini-conditions-mocca} outlines the initial conditions and code modifications applied to \textsc{MOCCA}, while Section~\ref{sec:mocca-results} presents the results on VMS growth and IMBH formation in the \textsc{MOCCA} simulation.

\section{Initial conditions}\label{sec:initialcond}
In this section we present both the star cluster parameters as well as the parameters employed in the modelling of stellar evolution.

\subsection{Star cluster parameters}
\label{sec:Star cluster parameters}

\begin{table}[h]
\caption[]{\label{table:IC}Initial conditions of our star cluster model and most important features related to the stellar evolution of all the stars.} 
\begin{tabular}{lcc}
\hline \hline
Parameter & Value & Ref. \\ \hline
Half-mass radius ($r_{\rm hm}$) &  $0.10~\mathrm{pc}$  &   \\
Half-mass average density ($\rho_{\rm hm}$) & $6.99\cdot 10^7 ~\rm M_\odot/{\rm pc}^3$ & \\
Density profile King ($W_0$)     &  $6$  &  (1) \\
Total particle number ($N$) &    $ 10^6 $  & \\
Absolute metallicity ($Z$)      &   $0.01$    &  \\
IMF, min/max stellar mass &  $0.08 - 150 ~\rm M_\odot$   &  (2) \\
Initial cluster mass ($M_{\rm i}$)  &  $5.86 \cdot 10^5 ~\rm M_\odot$ &  \\
Tidal Field  & none & \\
\hline \hline
\end{tabular}
\tablebib{
(1)~\citet{King1966};
(2)~\citet{Kroupa2001}
}
\end{table}
\begin{figure}[b!]
  \centering \includegraphics[width=\columnwidth]{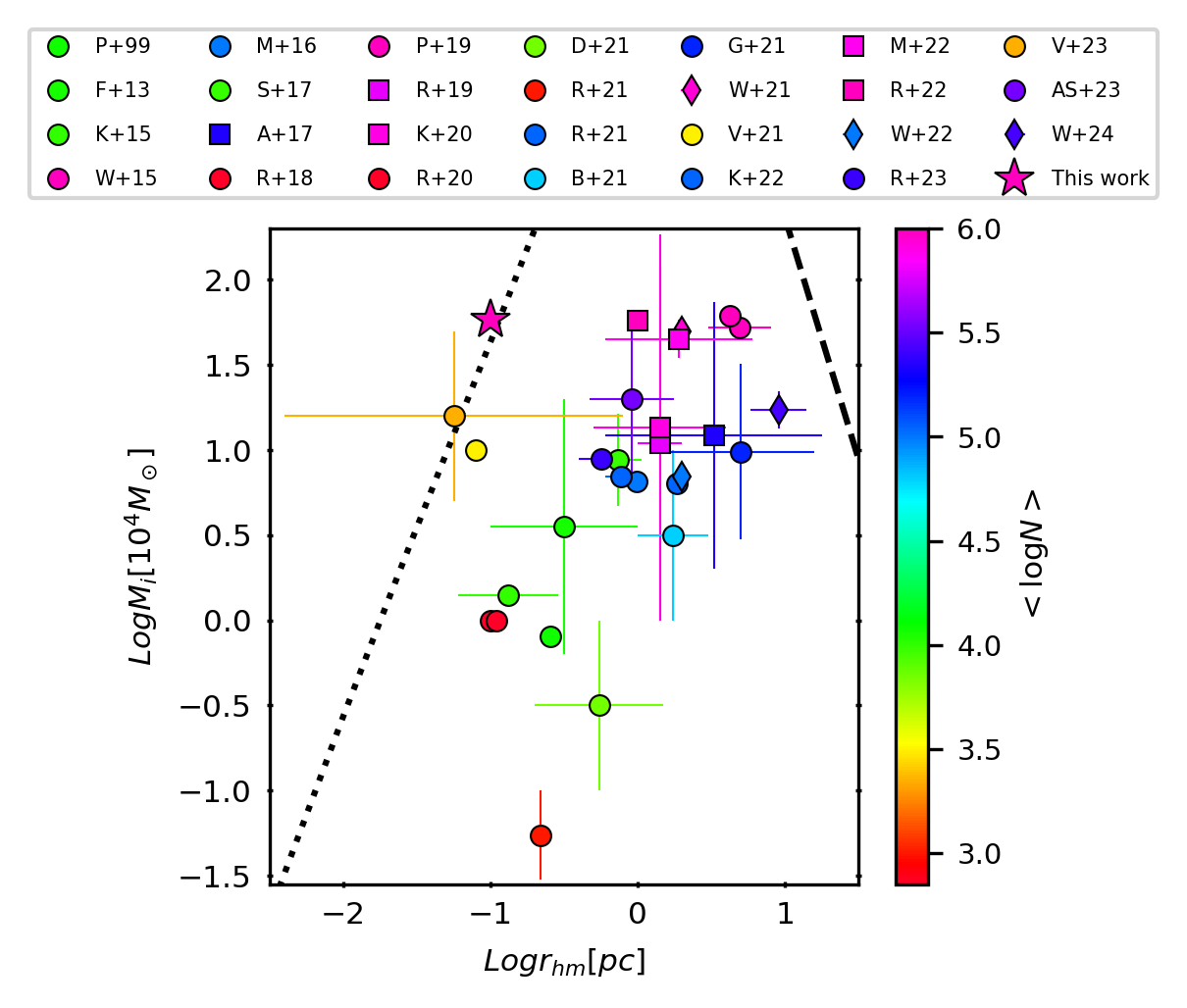}
  \caption{%
     Initial mass of the cluster as a function of the half-mass radius. The color bar represents the number of particles. The dashed black line represents the relaxation time and the dotted black line corresponds to the collision time, when both are equals to a time evolution (i.e. age) of $\tau = 10\,$Gyr, which we consider as an upper limit for the typical evolutionary times of the clusters. Symbols indicate different simulations: \textit{N}-body (squares), Monte Carlo (circles), hybrid \textit{N}-body (diamonds), and this work (star).
  }
  \label{fig:Mi_vs_rh_colorbar_N}
\end{figure}

In this paper we present one selected star cluster model, all relevant global parameters are given in Table~\ref{table:IC}. The dynamical evolution of this system is followed by \textsc{Nbody6++GPU} as well as \textsc{MOCCA}, for the purpose of comparison. The initial model was generated by \textsc{Mcluster} \citep{Kuepper2011, Leveque2021}. The initial number of stars is set to $10^6$, with no initial (primordial) binaries. While such binaries in principle can be included \citep[see e.g.][]{Wang2016,Rizzuto2021} they are computationally very expensive and also many free parameters for their initial configuration are added. We postpone models with many primordial binaries and a more extended parameter study of initially highly compact star cluster models to future work; in our simulations, however, binaries form and evolve dynamically.
The cluster density is modelled via a \citet{King1962} profile with central dimensionless well $W_0 = 6$. Initial stellar masses are distributed according to a \citet{Kroupa2001}  initial mass function limited between $0.08 - 150 ~\rm M_\odot$. All details about our cluster model are summarised in Table~\ref{table:IC}.
The cluster model is initially in virial equilibrium, with no initial mass-segregation, rotation , or any level of fractality \citep{Goodwin2004}. We do not use an external tidal field i.e., an external gravitational field from the host galaxy, since we are more interested in the dynamical evolution near and inside the half-mass radius. Still, it is possible that stars escape, if their total energy is positive and their distance from the cluster center is larger than 20 initial half-mass radii \cite{Aarseth2003a}.\\

Crossing time and relaxation time taken at the half-mass radius of a star cluster ($t_{\rm cr,hm}$ $t_{\rm rx,hm}$) are important parameters determining the general evolution of a star cluster \citep{Binney1987, Binney2008}. The crossing time corresponds to the time in which a star with a typical velocity travels through the cluster under the assumption of virial equilibrium. The half-mass relaxation time corresponds to the time over which the cumulative effect of stellar encounters on the velocity of a star becomes comparable to the star’s initial velocity. In our star cluster model the initial relaxation time is $7.01\,$ Myr, and the initial crossing time is $5.9 \cdot \ 10^{-4}\,$ Myr.

The mass segregation timescale is important to describe the initial segregation of our star cluster model \citep{Kroupa1995}. The mass segregation timescale is defined as:
\begin{equation}
    t_{\rm ms} = (\overline{m}/m_{\rm max})\, t_{\rm rx,hm}\,,
\end{equation}
where $\overline{m}$ is the initial average mass of the stars in the cluster, while $m_{\rm max}$ is the largest star mass in the cluster, which are 0.58 $~\rm M_\odot$ and 148.49 $~\rm M_\odot$, respectively. This, implies a segregation time of $t_{\rm ms} = 0.03$ Myr. Moreover, given the large density of the cluster model, this means that encounters between large masses occurs relatively early in the star cluster dynamical history.

Another relevant timescale to describe the stellar dynamics is the (average) collision time, which quantifies the rate of collisions between stars in the evolution  cluster \citep{Binney1987, Binney2008} is $t_{\rm coll} = 5.35 \cdot \ 10^{-3}\,$ Myr. According to the scenario proposed by \cite{Escala2021}, from the comparison of the relaxation and collision times to the age of the system $\tau$, it is possible to determine if collisions will play an important role in building a massive central object and the long term stability  of the cluster. To test this, \citet{Vergara2023} performed idealized (equal mass) numerical simulations and in order to quantify this transition, defined a critical mass for $t_{\rm coll}=\tau$, which can be expressed as follows:

\begin{equation} \label{eq_mass_crit}
M_{\mathrm{crit}} = r_{\mathrm{hm}}^{7/3} \left( \tfrac{4}{3} \pi \, m_* \, \Sigma_0^{-1} \tau^{-1} \sqrt{G} \right)^{2/3}\,,
\end{equation} 
where the effective cross-section is given by $\Sigma_0 = 16 \sqrt{\pi} R_*^2(1+ \Theta)$, the Safronov number is defined as $\Theta = 9.54 \left(M_* R_\odot/~\rm M_\odot R_*\right) \left(100~\rm{km\,s^{-1}}/\sigma\right)^2$, and $n$ is the numeric density. \cite{Vergara2023} found that the fraction of the total mass in the central object dramatically increase for $t_{\rm coll}<\tau$, being  up to 50\% for the most extreme runs.
 
Fig.~\ref{fig:Mi_vs_rh_colorbar_N}, shows the initial mass of the cluster against the half-mass radius and includes a color bar with the average logarithm of the number of stars. It also displays the relaxation time  and the collision time (or critical mass), when both are equals to a time evolution (i.e. age) of $\tau = 10\,$ Gyr, which we adopt here as an upper limit for the  evolutionary timescales in the simulations we are comparing with. We compare the respective simulations for several grids of direct $N$-body  shown as circle symbols in plot representing initial conditions from P+99: \citet{PortegiesZwart1999}, F+13: \citet{Fujii2013}, K+15: \citet{Katz2015}, W+15: \citet{Wang2015}, M+16: \citet{Mapelli2016}, S+17: \citet{Sakurai2017}, R+18: \citet{Reinoso2018}, P+19: \citet{Panamarev2019},  R+20: \citet{Reinoso2020}, D+20: \citet{DiCarlo2020a}, R+20: \citet{Rastello2021}, R+21: \citet{Rizzuto2021}, B+21: \citet{Banerjee2021}, G+21: \citet{Gieles2021},  V+21: \citet{Vergara2021}, K+22: \citet{Kamlah2022b}, R+23: \citet{Rizzuto2023}, V+23: \citet{Vergara2023}, and AS+23: \citet{ArcaSedda2023a, ArcaSedda2024a, ArcaSedda2024}, Monte-Carlo shown as square symbols in plot representing initial conditions from A+17: \citet{Askar2017a}, R+19: \citet{Rodriguez2019}, K+20: \citet{Kremer2020a}, M+22: \citet{Maliszewski2022}, and R+22: \citet{Rodriguez2022} and hybrid simulations diamond symbols in plot representing initial conditions from W+21: \citet{Wang2021}, W+22: \citet{Wang2022a}, and W24: \citet{Wang2024}. The latter are classified as hybrid $N$-body simulations, because they combine particle-particle (e.g. direct $N$-body) and particle-tree (e.g. Barnes-Hut tree) methods to speed-up calculations, and exploit the advantage of a highly parallelised implementation using the code \textsc{PeTaR} \citep{Wang2020a}. 

We would like to highlight that the simulations presented in this work breach into a completely uncharted region; the combination of initial particle number $N$ and the half-mass radius $r_{\mathrm{hm}} = 0.1$~pc. In particular, it presents the densest direct million body simulation ever published. 

\subsection{Initial Conditions for \textsc{MOCCA} and Code Modifications}\label{ini-conditions-mocca}

Similar to the direct \textit{N}-body runs, the input parameters that govern stellar and binary evolution prescriptions in \textsc{MOCCA} were set to \texttt{level C}  as described in \citet{Kamlah2022a} and Appendix~\ref{sec:evolution_parameters}. For these runs, the overall time step in \textsc{MOCCA} was reduced to 0.02 Myr. Given the short central relaxation time of this dense model, the smaller time step allowed us to better resolve the evolution and growth of the VMS. Additionally, it enabled more accurate capture of mass segregation and the effects of mass loss on the cluster's evolution. In the \textsc{MOCCA} run, three-body binary formation (3BBF) was specifically disabled for the VMS only when its mass exceeded $500 \ ~\rm M_\odot$ to allow for a more accurate comparison with \textit{N}-body results. This modification was necessary because, in \textsc{MOCCA}, a star in a binary is not allowed to undergo two-body hyperbolic collisions. If the VMS is in a binary, it can only grow through collisions or mergers during binary-single and binary-binary encounters. With 3BBF enabled for the VMS, it frequently formed a binary and remained bound for extended periods, preventing it from growing through two-body collisions. As a result, in the simulation with 3BBF enabled, the VMS reached a mass of only 35,000 \( \rm M_{\odot} \) within 4.5 Myr, significantly limiting its growth compared to the direct \textit{N}-body simulation. This restriction is a limitation of \textsc{MOCCA} that does not exist in \textsc{NBODY6++GPU}, where two-body collisions can still occur even if the VMS is part of a binary system. By disabling 3BBF for the VMS in \textsc{MOCCA}, we allowed it to undergo two-body collisions more freely, leading to a growth rate much closer to that observed in \textsc{NBODY6++GPU}. This adjustment was crucial for achieving a more consistent comparison between the two simulation methods. Recent work by \citet{Atallah2024} on 3BBF showed that in interactions with a large mass ratio, 3BBF favors pairing the two least massive bodies. Given these findings, disabling 3BBF for the VMS in \textsc{MOCCA} may not be an unreasonable approach, as it prevents the VMS from being preferentially locked into a binary with a much lower-mass companion, thereby allowing for more efficient growth through collisions.

%
\section{Results}
\label{sec:Results}

In this section, we present and discuss the evolution of the star cluster, the dynamics of the VMS, its evolution due to collisions, their impact on rejuvenation and the formation of an IMBH.

\subsection{Dynamical star cluster evolution and the evolution of a VMS and formation of an IMBH}
\label{sec:Global, dynamical star cluster evolution}

We dynamically evolve our system for $5\,$Myr, when already a VMS had formed and evolved into an IMBH. However, relaxation processes are still relevant, since mass segregation takes place much faster than the global half-mass relaxation time given above, with a difference of three order of magnitude \citep{Khalisi2007}.

\begin{figure}[th!]
  \centering
  \includegraphics[width=\columnwidth]{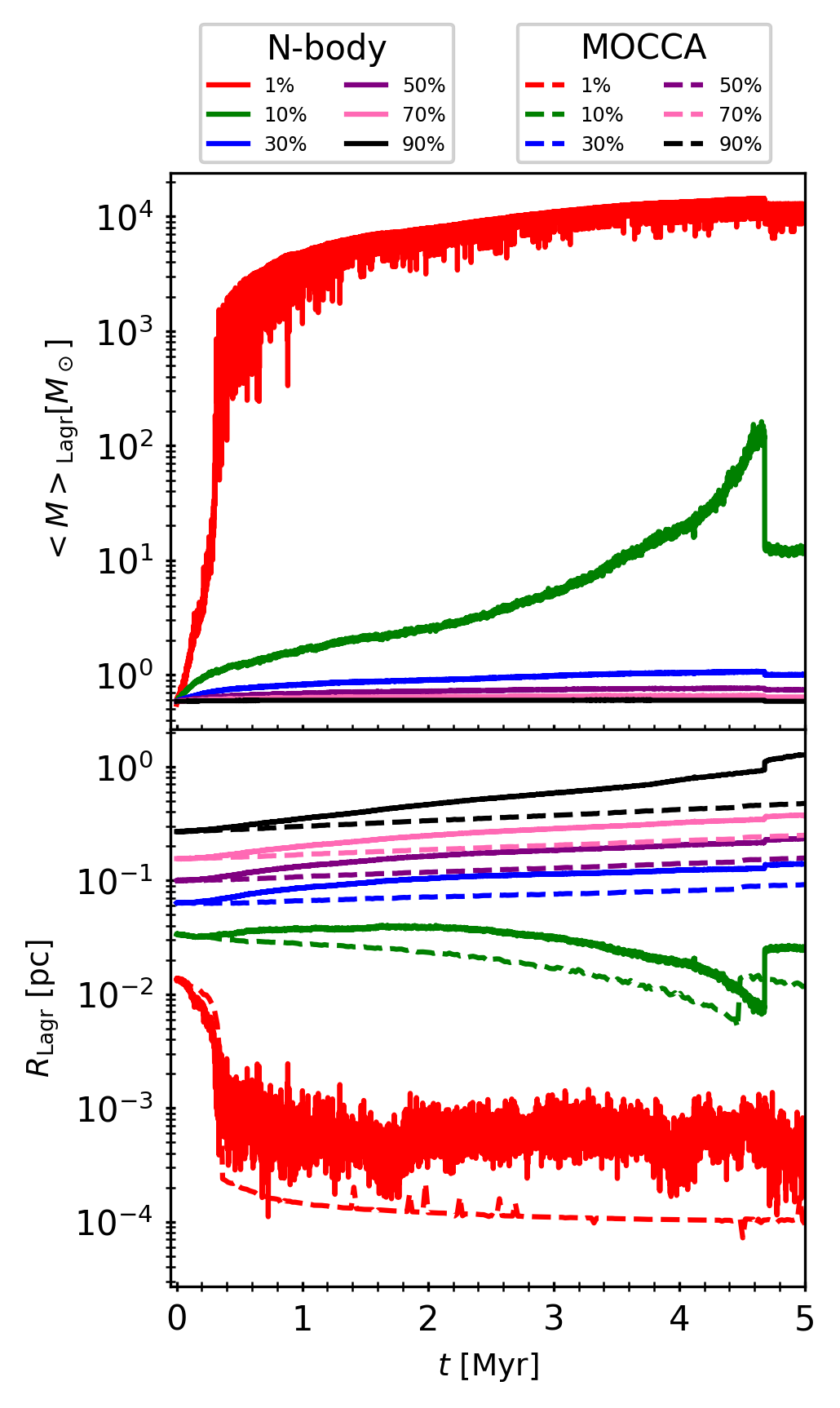}
  \caption{%
The figure shows the average mass in each shown Lagrangian shells $\langle M\rangle_{\mathrm{Lagr}}[\mathrm{M}_{\odot}]$ (Top) and, the Lagrangian radii $R_{\mathrm{Lagr}}~[\mathrm{pc}]$ (Bottom), both normalised by the cluster mass at the beginning of the simulation. We show both \textit{N}-body and \textsc{MOCCA} results for the Lagrangian radii. In both plots, we show the shell representing 1 \%, 10 \%, 30\%, 50\%, 70\% and 90 \%. The differences between the Lagrangian radii in the \textit{N}-body and \textsc{MOCCA} simulations arise from how escaping stars are treated. In the \textit{N}-body simulations, stars with positive energy are not immediately removed from the system; instead, they remain until they cross a boundary set at ten times the half-mass radius. These stars are still included in the calculation of Lagrangian radii, which affects the results compared to \textsc{MOCCA}.
} 
  \label{fig:rh01_global_final}
\end{figure}
Figure~\ref{fig:rh01_global_final} shows the average mass of the cluster inside the Lagrangian radii (top) and the Lagrangian radii (bottom), which shows the different evolution of shells from the inner to the outer regions. In our model, the star cluster evolves rapidly in the innermost regions ($1\%$), where strong mass segregation triggers an early core collapse. The $10\%$ Lagragian radii start to decrease after $3\,$Myr of evolution. The $30\%$, $50\%$, $70\%$ and $90\%$ Lagrangian radii show the smooth expansion of the cluster; Until around $4.69\,$Myr (solid lines) and $4.48\,$Myr (dashed lines) when the IMBH is formed, in \textit{N}-body and \textsc{MOCCA} simulations, respectively. 
For the innermost mass shell (1\%) we can follow the formation of the VMS, as its mass is getting larger than the corresponding Lagrangian shell's mass. Notably, the VMS reaches a mass of $M_{\rm VMS} \geq 1000 ~\rm M_\odot$ at time $0.18\,$Myr (\textit{N}-body simulation) and $0.22\,$Myr (\textsc{MOCCA} simulation).

\begin{figure}[bh!]
  \centering
  \includegraphics[width=\columnwidth]{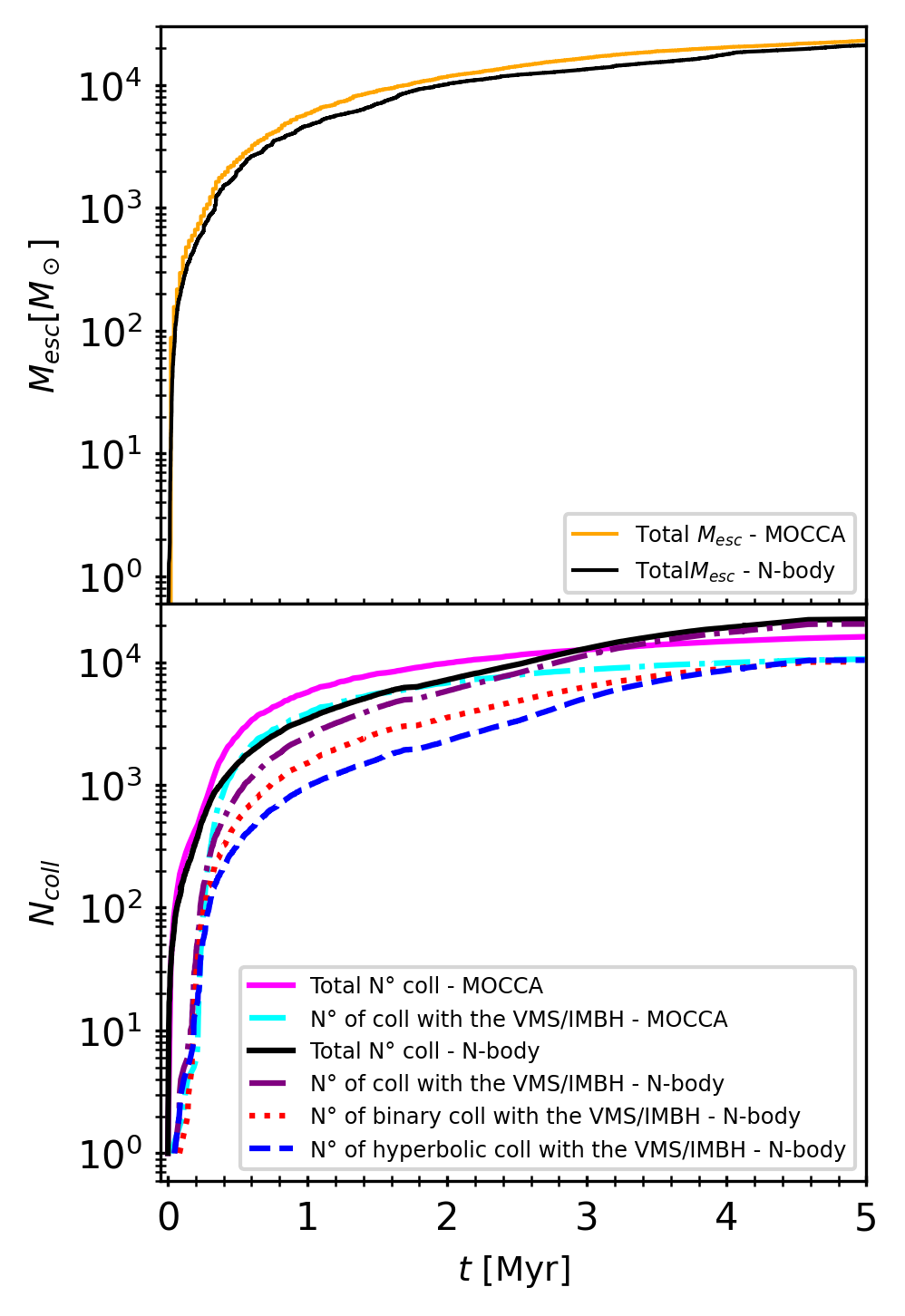}
  \caption{%
The Figure shows the cumulative mass of escapers in the top panel and the cumulative number of collisions in the bottom panel. The total number of collisions is represented by a black solid line (\textit{N}-body) and magenta solid line (\textsc{MOCCA}), while collisions involving the VMS are shown with a purple dot-dashed line (\textit{N}-body) and cyan solid line (\textsc{MOCCA}). Binaries involving the VMS are shown with a red dotted line, and hyperbolic collisions with the VMS are shown by a blue dashed line.
} 
  \label{fig:rh01_esc_coll}
\end{figure}

Figure~\ref{fig:rh01_esc_coll} shows the cumulative mass of the escapers during the simulation. In the first Myrs, the cumulative escaped mass is approximately $4\,000-5\,000~\rm M_\odot$, increasing to around $20\,000~\rm M_\odot$ after $5\,$Myr. \textit{N}-body and \textsc{MOCCA} simulation show reasonable agreement. The figure also presents the total number of collisions, which amounts to $22\,010$. Of these, $91.6\%$ involves the VMS/IMBH, these collisions can be divided into binary collisions ($45.1\%$) and hyperbolic collisions ($46.5\%$). Binary collisions are more frequent when the VMS mass exceeds $1\,000~\rm M_\odot$ at time $0.18\,$Myr. In contrast, hyperbolic collisions become more frequent as the VMS grows in size, reaching a radius larger than $1\,000~\rm R_\odot$ at time $3.36\,$Myr.
For the \textsc{MOCCA} simulation, we do not distinguish between binary and hyperbolic collisions, as the code does not explicitly track stellar orbits. Some collisions may involve stars that were gravitationally bound to the VMS, but confirming this would require reconstructing stellar orbits from snapshots using their instantaneous positions, velocities, and the local gravitational potential, which is challenging due to the limited temporal resolution. Consequently, we report only the total number of collisions ($16103$ in 5 Myr), of which $65.6\%$ involve the VMS/IMBH. Despite this limitation, the results from both codes are in good agreement.

\begin{figure*}
  \centering
  \includegraphics[width=\textwidth]{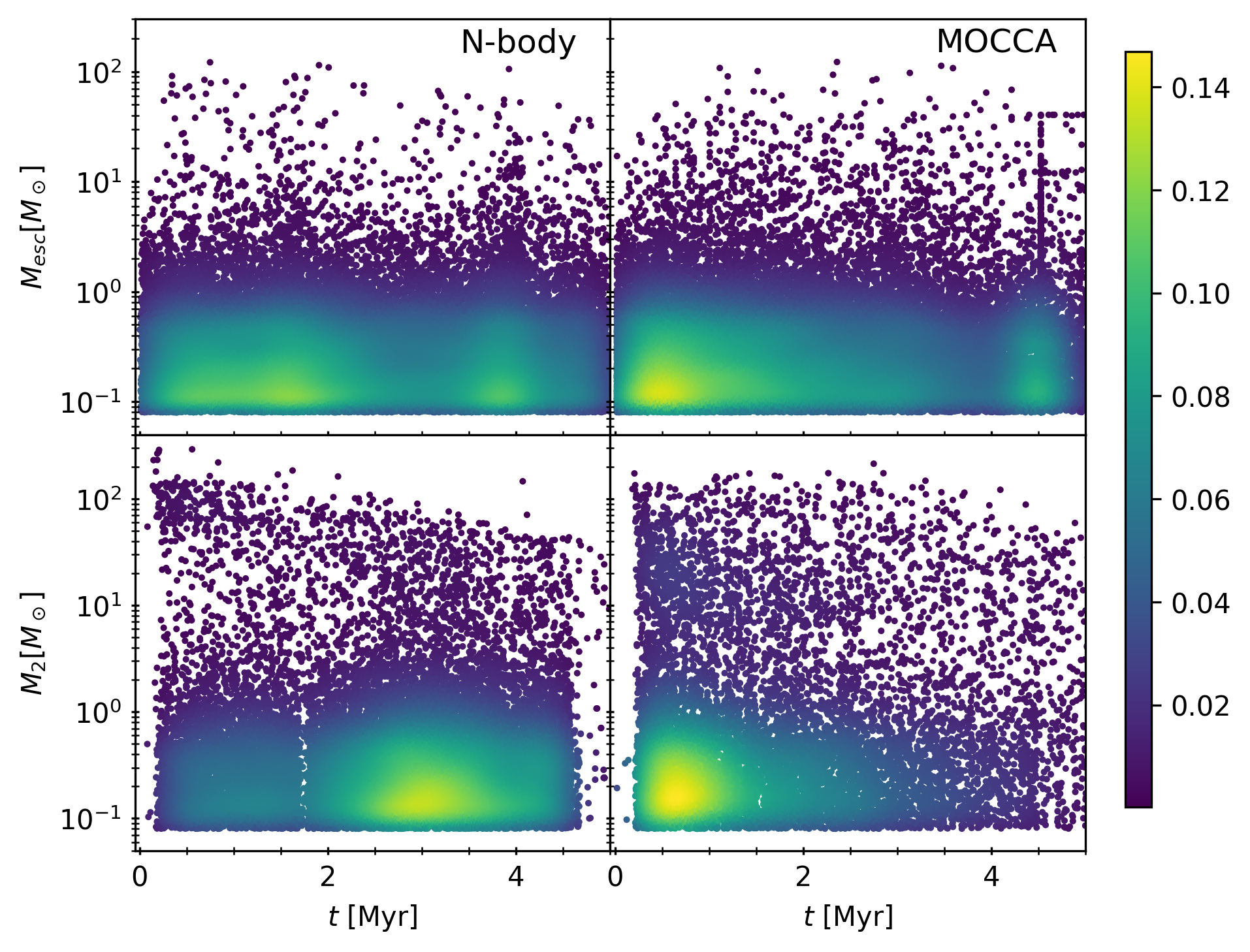}
  \caption{%
    The Figure shows, in top panels: the mass of each escaping star. In bottom panels: the mass of the secondary star before colliding with the VMS, both in the case of hyperbolic and binary collisions. The color bar represents the density distribution of stars
  }
  \label{fig:escapers_secondary_mass}
\end{figure*}

Figure~\ref{fig:escapers_secondary_mass} compares the \textsc{NBODY6++GPU} and \textsc{MOCCA} simulations, showing the time evolution of the masses of escapers from the cluster (top panels) and secondary stars involved in collisions with the VMS/IMBH (bottom panels). The left panels display the \textit{N}-body results, and the right panels show the \textsc{MOCCA} results. The color scale represents the local point density of events, with yellow regions indicating higher concentrations of recorded data points. In both simulations, low-mass stars (typically $\lesssim 1~\rm M_\odot$) dominate the population of escapers. The \textit{N}-body simulation shows two prominent epochs of escaper activity: an extended early phase over the first $\sim2\,$Myr, followed by a shorter but more intense burst near the time of IMBH formation at $4.69\,$Myr. In contrast, the \textsc{MOCCA} simulation exhibits a smoother and more gradual decline in escaper activity over time, with a modest peak near $4.48$ Myr corresponding to IMBH formation.

\begin{figure*}
  \centering
  \includegraphics[width=\textwidth]{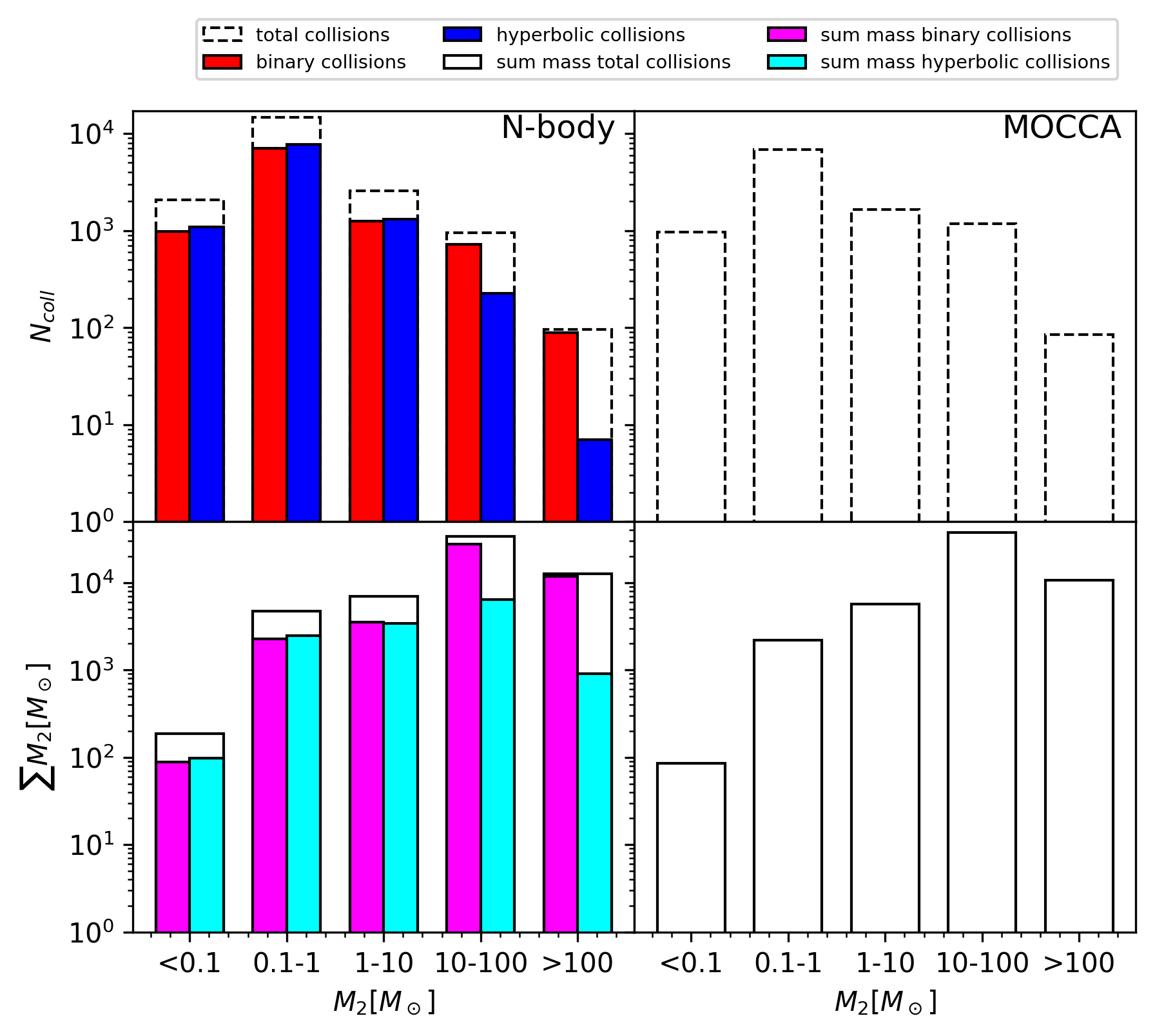}
  \caption{%
   Histograms showing the contribution of colliding stars to VMS formation, divided into binary, hyperbolic, and total collisions. The mass ranges displayed are: $<0.1~\mathrm{M}_\odot$, $0.1$–$1~\mathrm{M}_\odot$, $1$–$10~\mathrm{M}_\odot$, $10$–$100~\mathrm{M}_\odot$, and $>100~\mathrm{M}_\odot$.
Results are shown for both the \textit{N}-body simulation (left column) and the \textsc{MOCCA} simulation (right column).
The top panel displays the number of collisions, while the bottom panel shows the cumulative mass contributed by these collisions. }
  \label{fig:histogram_bin_n_hyp}
\end{figure*}

The lower panels show the masses of stars colliding with the VMS/IMBH. \textsc{MOCCA} displays a broader and more uniform spread across the stellar mass range from the start of the simulation, while the \textit{N}-body simulation shows occasional high-mass collisions (with $m \gtrsim 10~\rm M_\odot$), though these are relatively sparse. In both simulations, most collisions involve low-mass stars. However, these frequent low-mass mergers contribute less significantly to the final VMS/IMBH mass (see also Figure~\ref{fig:histogram_bin_n_hyp}). The \textit{N}-body simulation shows a brief dip in the number of secondary stars involved in collisions around $1.68\,$Myr, lasting approximately $0.1\,$Myr. This feature coincides with the formation of a hard binary between the VMS and a main-sequence star (see Figure~\ref{fig:hard_binary}). Subsequent three-body interactions lead to enhanced escape rates. Toward the end of the simulation, the IMBH sequentially forms two hard binaries with $\sim2~\rm M_\odot$ MS stars, which explains the observed drop in further collisions.

Figure ~\ref{fig:histogram_bin_n_hyp} shows histograms of the number of collisions (top panels) and the cumulative mass of the secondary star colliding with the VMS/IMBH (bottom panels). The \textit{N}-body results are shown in the left panels, while the \textsc{MOCCA} results are shown in the right panels, the histogram mass ranges $<0.1$, $0.1-1$, $1-10$, $10-100$ and $>100~\rm M_\odot$. The number of collisions for \textit{N}-body are $2057$, $14525$, $2525$, $923$, $95$, while for \textsc{MOCCA} are $1053$, $7797$, $2273$, $1570$, $74$, for the first two mass ranges, \textit{N}-body has about twice as many collisions as \textsc{MOCCA}, for the mass range $1-10~\rm M_\odot$ the number of collisions is about 300 more in \textit{N}-body, while for the next mass range $10-100~\rm M_\odot$ it is higher for \textsc{MOCCA}, about 600 more collisions. Finally, for the last mass range $>100~\rm M_\odot$ the number of collisions is slightly large in the \textit{N}-body simulation. In the case of \textit{N}-body simulations where it is possible to distinguish between binary (red bars) and hyperbolic (blue bars) collisions, we can observe that hyperbolic collisions contribute slightly more to the VMS formation for the mass range $<0.1$, $0.1-1~\rm M_\odot$. The mass range $1-10~\rm M_\odot$ shows almost the same number of collisions, however, binary collisions contribute slightly more to the VMS mass, while binary collisions contribute quite a bit more to the VMS formation for the mass range $10-100$, and $>100~\rm M_\odot$. Both codes shows the largest number of collisions comes from stars in the mass range $0.1-1~\rm M_\odot$. However, the largest mass contribution to the VMS comes from stars in the mass range $10-100~\rm M_\odot$, for \textsc{MOCCA} the mass contribution is $4.41 \times 10^4 ~\rm M_\odot$, while in \textit{N}-body it is $3.35 \times 10^4 ~\rm M_\odot$ ($2.73 \times 10^4 ~\rm M_\odot$ from binary collisions (cyan bar) and $6.34 \times 10^3 ~\rm M_\odot$ from hyperbolic collisions (magenta bar)), the second large mass contribution to the VMS comes from the mass range $>100~\rm M_\odot$, in \textsc{MOCCA} with $9.46 \times 10^3 ~\rm M_\odot$ and in \textit{N}-body with $1.26 \times 10^4 ~\rm M_\odot$, with $1.17 \times 10^4$ and $914.19 ~\rm M_\odot$, from binary and hyperbolic collisions, respectively. For the lower mass ranges $<0.1$, $0.1-1$ and, $1-10~\rm M_\odot$ the VMS mass contribution in the same order is $94.49$, $2.59 \times 10^3 ~\rm M_\odot$ and, $8.18 \times 10^3 ~\rm M_\odot$ for \textsc{MOCCA}, while for \textit{N}-body it is $184.01$, $4.69 \times 10^3 ~\rm M_\odot$ and, $6.80 \times 10^3 ~\rm M_\odot$, for this mass ranges the VMS mass contribution from binary and hyperbolic collisions is quite similar.

\subsection{N-body results: VMS evolution and IMBH formation}
\label{sec:Nbody result}
In our \textit{N}-body simulation, several collisions along the simulation to build up a VMS with a mass of $56996.91~\rm M_\odot$, eventually collapsing and forming an IMBH with a mass of $51625.99~\rm M_\odot$ at time $4.69\,$Myr.

\begin{figure}
  \centering
  \includegraphics[width=\columnwidth]{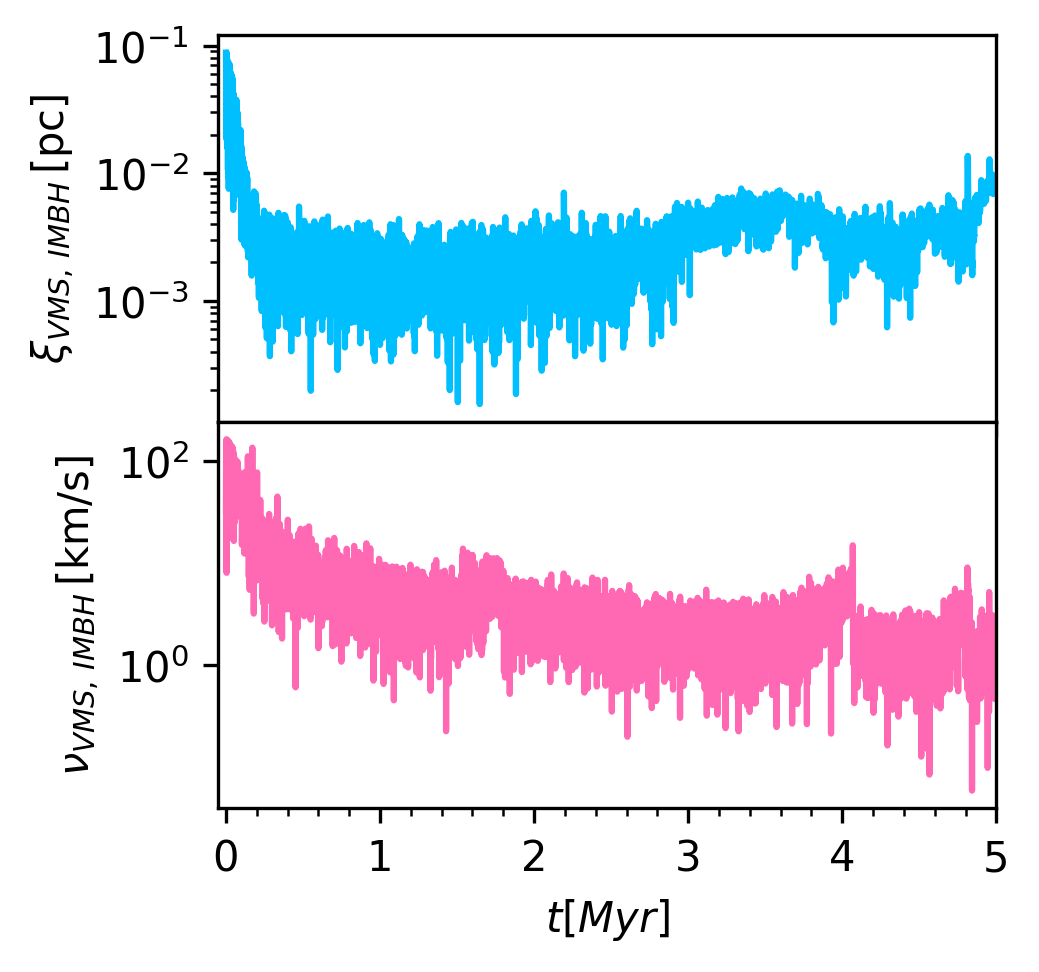}
  \caption{%
The figure shows the temporal evolution of the position of the VMS relative to the star cluster density center, $\xi_{\rm VMS\,IMBH}$, and the speed of the VMS relative to the star cluster density center, $\nu_{\rm VMS\,IMBH}$.}
  \label{fig:vms_pos_vel}
\end{figure}

\begin{figure}[b!]
  \centering
  \includegraphics[width=\columnwidth]{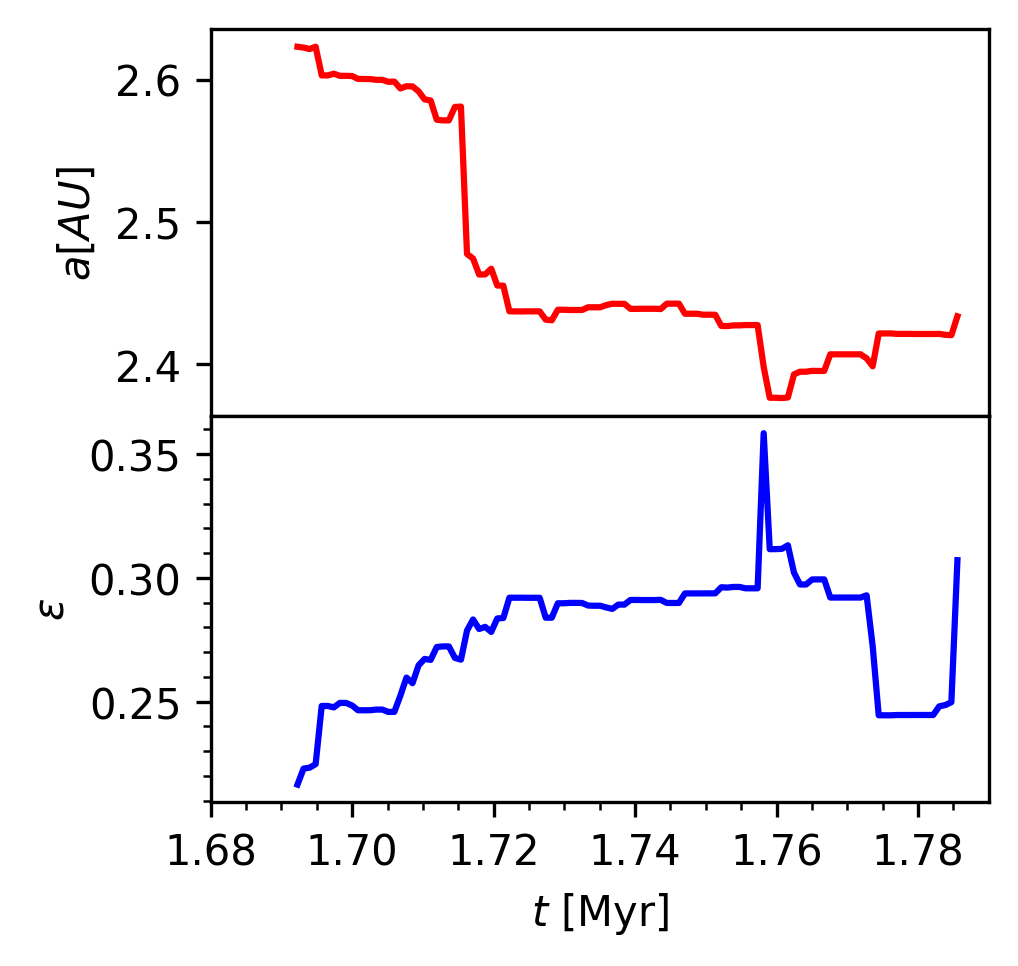}
  \caption{%
  The Figure shows the temporal evolution of a hard binary system during between 1.68 and 1.80 Myr. This binary has the primary star as the VMS, with a mass of $28891~\rm M_\odot$ and a secondary star with a mass of $75~\rm M_\odot$. This interaction lasts approximately $0.1$ Myr. The top panel shows the semi-major axis, while the bottom panel shows the eccentricity.}  
  \label{fig:hard_binary}
\end{figure}

Figure~\ref{fig:vms_pos_vel} shows the scatter of the position relative to the star cluster's density center of the VMS (and IMBH thereafter), $\xi_{\mathrm{VMS,\,IMBH}}~\rm[pc]$, and the speed of the VMS (and IMBH, when formed), $\nu_{\mathrm{VMS,\,IMBH}}~[\mathrm{km\,s}^{-1}]$. Initially, the star with a mass of $130.58~\rm M_\odot$, which later becomes the VMS, is located at $0.088\rm pc$ from the center, sinking with a velocity of $32.9~\rm{km\,s}^{-1}$. After around 100 years, it moves closer to the center at $0.061\rm pc$, while its velocity increases nearly threefold to $99.7~\rm{km\,s}^{-1}$. A few hundreds years later, it reaches a velocity of $165.1~\rm{km\,s}^{-1}$ at a position of $0.018\rm pc$. This rapid infall toward the center is driven by the high cluster density, however, the VMS position and velocity oscillations are consistent with the orbital time. After two collisions with MS stars (i.e., stars with a fraction of a solar mass), the first significant collision occurs with a star of mass $54.89 ~\rm M_\odot$ following a binary interaction. At this point, the star is at $0.006\rm pc$. As shown in the figure, the motion of the VMS, in terms of position and velocity, first aligns with the expectations of mass segregation (steady decrease in distance to the center and a velocity profile that initially rises before declining) and later transitions to Brownian motion. At $1.68\,$Myr a hard binary system forms between the VMS and an MS star. Figure~\ref{fig:hard_binary} shows the semi-mayor axis and eccentricity of the hard binary formed between the VMS with a mass of $28\,891~\rm M_\odot$ and a secondary star with a mass of $75~\rm M_\odot$. Both stars are tightly bound due to their small average distance ($< 3$ AU), they have an elliptical orbit, this dynamical interaction lasted about $0.1\,$Myr. The binary system have a higher binding energy than the kinetic energy of the surrounding stars, which eventually gain kinetic energy, increasing their velocity. The latter reduces the probability of low-velocity stars that could collide with the VMS. Throughout the simulation, the velocity and position of the VMS remain nearly constant, fluctuating around an average value of the order of a $10^{-3}\rm pc$, while the velocity remains in moderate motion, at a few kilometers per second. 

\begin{figure*}
  \centering
  \includegraphics[width=\textwidth, height=0.6\textheight]{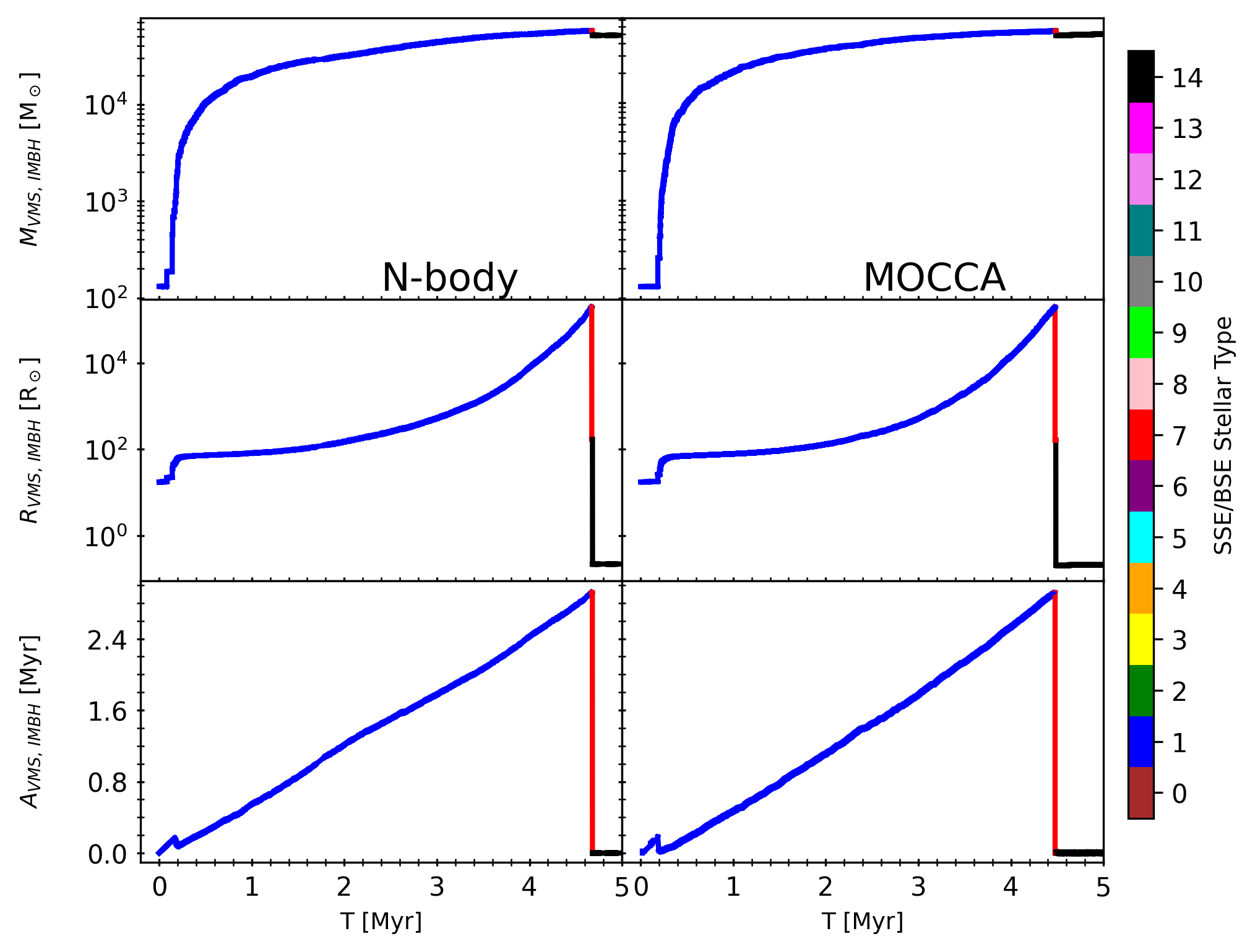}
   \caption{%
Evolution of the VMS (and IMBH thereafter) over time. 
Top panel: Mass of the VMS/IMBH, $M_{\mathrm{VMS\,IMBH}}~[\mathrm{M}_{\odot}]$. 
Middle panel: Stellar radius of the VMS/IMBH, $R_{\mathrm{VMS\,IMBH}}~[\mathrm{R}_{\odot}]$ (logarithmic scale). 
Bottom panel: Effective age of the VMS during its main-sequence phase, $A_{\mathrm{VMS\,IMBH}}~[\mathrm{Myr}]$.
}

  \label{fig:VMS_M_Rsun_ageeffective.jpg}
\end{figure*}

As supported by the theoretical expectations discussed in previous sections, the VMS growth in the \textit{N}-body simulation occurs on a short time-scale. The VMS original progenitor passes from an initial mass of $130.58~\rm M_\odot$ to exceed $10^3~\rm M_\odot$ after only $0.18\,$Myr through 17 collisions, further reading $>5000~\rm M_\odot$ in a time $0.29\,$Myr. 
The VMS continues its growth for $0.48\,$Myr through several collisions until reaching more than $10^4 ~\rm M_\odot$, naturally, the size of the VMS also increases having $71.13\rm R_\odot$, the continued collisions allow a strong rejuvenation of the star which has an effective age $0.23\,$Myr. Lately, around $1.0\,$Myr the VMS has a mass, effective age, and radius of $1.9 \times 10^4~\rm M_\odot$, $0.54\,$Myr, and $80.75\rm R_\odot$, respectively. The evolution of the mass, radius, and age of the VMS is shown in the top, middle, and bottom left panels of Figure~\ref{fig:VMS_M_Rsun_ageeffective.jpg}.

At $2\,$Myr, the VMS has a mass of $3.1 \times 10^4~\rm M_\odot$, with a radius $146.24\rm R_\odot$ and effective age $1.21\,$Myr, the hard binary system formed at $1.68\,$Myr in the \textit{N}-body simulation prevents the VMS from experiencing further collisions, thus leads to a VMS with about $4\,000 ~\rm M_\odot$ less mass than the VMS obtained in \textsc{MOCCA} simulation, that has a mass  of $3.5 \times 10^4~\rm M_\odot$ at the same simulation time, $1\,$Myr later the \textit{N}-body VMS reaches a mass, radius, and effective age of $4.3 \times 10^4~\rm M_\odot$, $507.24\rm R_\odot$ and $1.77\,$Myr, respectively. The \textit{N}-body VMS still have less mass than the \textsc{MOCCA} VMS which have a mass $4.52\times10^4~\rm M_\odot$ at $3\,$Myr. 

The hydrogen burning phase is set to end in $2.9\,$Myr, according to the SSE/BSE prescription \cite{Hurley2000, Hurley2002a, Hurley2005}. After the VMS reaches this point, it should start burning helium. At $4.67\,$Myr the VMS has a mass of $5.7362\times10^4~\rm M_\odot$, with a radius of $1.9\times10^5\rm R_\odot$ and effective age of $2.91\,$Myr. At this stage, the VMS becomes a helium main sequence (HeMS; or type 7 in SSE/BSE by \cite{Hurley2000, Hurley2002a, Hurley2005})) star with a radius of $171\rm R_\odot$. During this helium phase, the VMS experiences three collisions with MS stars within $0.02\,$Myr before naturally collapsing into an IMBH. This process may result in mass loss of about $10~\%$ due to neutrino emission \citep{Fryer2012, Kamlah2022a}, leading to a mass of $51\,625.99~\rm M_\odot$ at time $4.69\,$Myr. Finally, the IMBH undergoes additional collisions, accreting $50\%$ of the stellar companion \citep{Rizzuto2021}, reaching a final mass of $51\,661.21~\rm M_\odot$ by the end of the simulation.

\subsection{MOCCA results: VMS evolution and IMBH formation}
\label{sec:mocca-results}

In the \textsc{MOCCA} simulation, the star that eventually grows into a VMS has an initial mass of $\rm 130.6 \ ~\rm M_\odot$ and is initially located at a distance of 0.088 pc from the center of the cluster. Within 0.15 Myr, this massive star undergoes four collisions with low mass stars, slightly increasing its mass to approximately $\rm 132 \ ~\rm M_\odot$. 
For a brief period between about $0.18$ Myr and $0.21$ Myr, the VMS is part of a binary system. At $0.182$ Myr, the VMS first forms a binary with a $\rm 72.4 \, M_\odot$ star due to 3BBF. Shortly after, at $0.183$ Myr, this companion is replaced by a $\rm 124 \, M_\odot$ star in an exchange encounter. Following several flyby interactions, the VMS and its binary companion merge during a binary-single interaction at $0.186$ Myr. The VMS subsequently forms another binary, first with a $\rm 72.4 \, M_\odot$ star, then with a $\rm 68 \, M_\odot$ star, and later with a $\rm 174 \, M_\odot$ star. The latter also merges with the VMS in a binary-single interaction at around $0.21$ Myr, following several flyby encounters. At this stage, the VMS is located at a distance of approximately $3 \times 10^{-4} \ \rm pc$ from the cluster center.
By $0.212$ Myr, its mass exceeds $\rm 500 \, M_\odot$ following a collision with a $\rm 121.8 \, M_\odot$ star, with which it was also briefly in a binary system. As described in Section~\ref{ini-conditions-mocca}, from this point onward, we disable 3BBF for the VMS. Its mass continues to increase rapidly as it undergoes runaway collisions with other MS stars in the cluster center. By 0.23 Myr, the VMS mass reaches $\rm 1000 \ ~\rm M_\odot$, and by 0.34 Myr, it exceeds $\rm 5000 \ ~\rm M_\odot$. The growth of the VMS in \textsc{MOCCA} is shown in the top-right panel of Fig. \ref{fig:VMS_M_Rsun_ageeffective.jpg}.

During these early stages of VMS growth, collisions with other VMSs also strongly rejuvenate it. At 0.34 Myr, when its mass reaches $\rm 5000 \ ~\rm M_\odot$, its effective age is only 0.068 Myr, and its radius is approximately $\rm 68 \ \rsol$. By 0.51 Myr, the VMS mass surpasses $\rm 10^{4} \ ~\rm M_\odot$, at which point its effective age is 0.17 Myr, and its radius is $\rm 71.5 \ \rsol$. The VMS continues its rapid growth, reaching $\rm 2.1 \times 10^{4} \ ~\rm M_\odot$ by 1 Myr. By this time, the cumulative number of collisions involving the VMS is approximately 3,815 (see Fig. \ref{fig:rh01_esc_coll}). As the mass ratio of these collisions decreases with increasing VMS mass, its effective rejuvenation per collision also weakens. At 1 Myr, its effective age is 0.468 Myr. The evolution of the VMS's effective age on the MS is shown in the bottom-right panel of Fig. \ref{fig:VMS_M_Rsun_ageeffective.jpg}.

By 2 Myr, the VMS reaches a mass of $\rm 3.56 \times 10^{4} \ ~\rm M_\odot$, with an effective age of 1.1 Myr and a radius of $\rm 131 \ \rsol$. Up until this point, the VMS has undergone more than 6,793 collisions. Between 2 Myr and 3 Myr, its growth slows, and by 3 Myr, it reaches a mass of $\rm 4.59 \times 10^{4} \ ~\rm M_\odot$ with an effective age of 1.77 Myr. This corresponds to approximately 61\% of the MS lifetime of the VMS, which is set to 2.91 Myr in the SSE/BSE prescriptions used in \textsc{NBODY6++GPU} and \textsc{MOCCA}. As the VMS ages, its radius increases, reaching $\rm 523 \ \rsol$ at 3 Myr. The evolution of the VMS's radius is shown in the middle-right panel of Fig. \ref{fig:VMS_M_Rsun_ageeffective.jpg}.

At 4.477 Myr, the VMS reaches a mass of $\rm 5.38 \times 10^{4} \ ~\rm M_\odot$, with a radius of $\rm 1.9 \times 10^{5} \rsol$ and an effective age of 2.91 Myr, marking the MS turn-off time for the VMS. By this point, it has undergone 10,322 collisions. After this stage, the VMS evolves into a helium main sequence star (HeMS; type 7 in SSE/BSE) with a radius of $\rm 165 \ \rsol$. Between 4.477 Myr and 4.485 Myr, the HeMS VMS undergoes approximately 15 additional collisions with lower-mass stars, reaching a final mass of $\rm 5.384 \times 10^{4} \ ~\rm M_\odot$. As the mass ratios in these collisions are extremely small, no rejuvenation is applied. At 4.485 Myr, the VMS evolves into an IMBH with a mass of $\rm 4.845 \times 10^{4} \ ~\rm M_\odot$. This IMBH then continues to grow through collisions, mostly with other stars, assuming that 50\% of the stellar mass is accreted by the IMBH in each collision. By 5 Myr, it reaches a mass of $\rm 4.984 \times 10^{4} \ ~\rm M_\odot$.

The short duration of the HeMS phase in the evolution of the VMS is a result of the inverse dependence of the HeMS lifetime ($t_{\mathrm{HeMS}}$) on stellar mass \citep[see Equation 79 in][]{Hurley2000}. As discussed in Section~\ref{sec:Nbody result}, the amount of mass lost due to neutrinos when the VMS transitions into an IMBH is set to 10\%, consistent with the default assumptions in \textsc{NBODY6++GPU}\footnote{In \textsc{MOCCA}, the default is to assume 1\% mass loss due to neutrinos when an evolved star becomes a BH}. We also performed an alternative \textsc{MOCCA} run in which the HeMS lifetime of the VMS was extended to 0.3 Myr. In this case, the VMS remained on the HeMS long enough to undergo a collision with a $\rm 53.3 \ ~\rm M_\odot$ stellar-mass BH present in the cluster. Although the mass ratio of this merger was extremely small, if the VMS was allowed to change its stellar type following the collision, the BH was treated as absorbing 50\%\footnote{The fraction of mass absorbed in collisions between stars and BHs is an input parameter specified in the initial conditions.} of the VMS mass. This alternative treatment led to the formation of an IMBH with a mass of $\rm 2.6725 \times 10^{4} \ ~\rm M_\odot$.

Overall, the \textsc{MOCCA} evolution of the VMS mass, radius, and effective age in the simulation presented here is in good agreement with the \textsc{NBODY6++GPU} results, as shown in Fig.\, \ref{fig:VMS_M_Rsun_ageeffective.jpg}. The mass distribution and number of stars that collide with the VMS are shown in Fig.\,\ref{fig:histogram_bin_n_hyp}, where the top right panel highlights that the total number of collisions involving the VMS is lower in \textsc{MOCCA} compared to the direct \textit{N}-body simulation (top left panel). This results in less efficient rejuvenation, allowing the VMS to reach the end of its main-sequence lifetime and evolve into an IMBH slightly earlier (by about 0.2 Myr) than in the \textit{N}-body run. Despite the reduced number of collisions, the final VMS mass at the time of IMBH formation is only $\rm 3500 \ ~\rm M_\odot$ smaller in \textsc{MOCCA}, due to the fact that no mass is lost during MS collisions in this simulation.

%

\section{Summary, conclusions and outlook}
\label{sec:summary}

This work investigates the formation of VMSs through stellar mergers, using both direct \textit{N}-body and \textsc{MOCCA} simulations. To better follow the evolution and growth of a VMS in these models, several improvements were made to the standard stellar evolution routines in \textsc{NBODY6++GPU} and \textsc{MOCCA}.
These included updates to the treatment of stellar radius evolution, rejuvenation, and close encounter prescriptions. We note that the prescriptions implemented here represent reasonable extrapolations of standard SSE/BSE stellar evolution models  \citep{Hurley2000, Hurley2002a, Hurley2005,Kamlah2022a}; however, the internal structure and thermal state of VMS formed through stellar collisions remain to be established and require further refinement. Our simulations demonstrate the growth of a VMS through successive mergers, reaching several tens of thousands of solar masses. The formation of such a massive object through a purely collisional pathway highlights the potential importance of this channel in the formation of IMBHs and SMBH seeds. This dual approach allows us to test the robustness of the VMS formation channel across methods with different computational schemes and resolution limits. 
We note that the \textsc{MOCCA} results required some adjustments to achieve consistency with the \textsc{NBODY6++GPU} simulations. In particular, the timestep was reduced to better capture the mass segregation time of massive stars. Without this change, a VMS would still have formed and reached a comparable final mass to the \textit{N}-body run, but its growth would be delayed by  $\approx$0.2 Myr. In addition, 3BBF was disabled for the VMS once it exceeded $500~\rm M_\odot$. With 3BBF enabled, the VMS quickly forms a binary and is prevented from undergoing two-body collisions, reaching only $\sim 35{,}000~\rm M_\odot$ at $4.5$ Myr. In contrast, with 3BBF disabled, the collisional growth of the VMS proceeds at a rate consistent with the direct \textit{N}-body result. These points are discussed in more detail in Subsection~\ref{ini-conditions-mocca}.

\subsection{Theoretical considerations and simulation constraints}\label{sec:summary-theory}

The formation of VMSs through runaway collisions in dense star clusters was first proposed by \citet{Begelman1978}, \citet{Rees1984}, and \citet{Lee1987} using simple analytical models. This scenario was later explored using Fokker–Planck and Monte Carlo methods by \citet{Quinlan1990, Gurkan2004, Freitag2006a, Freitag2006} and \citet{Sharma2025}, and more recently through direct \textit{N}-body simulations by \citet{Vergara2023}.
In clusters with a larger number of particles, dynamical equilibrium is typically reached before thermodynamic equilibrium. As a result, such systems require longer times to undergo core collapse and reach Spitzer instability without significant expansion due to relaxation processes \citep{Spitzer1987}. Our results support the existence of a critical mass threshold for the formation of massive objects via collisions, as previously suggested by \citet{Vergara2023, Vergara2024}. Our simulations push into an uncharted regime of extremely high central density and particle number, testing the runaway collision scenario under extreme but plausible cluster conditions.

The high central density of the clusters in our models ($>10^7~\rm M_\odot~\rm pc^{-3}$) leads to an elevated stellar collision rate, significantly higher than that in typical observed globular clusters. Consequently, the time between successive collisions becomes much shorter than the thermal timescale of the VMS. This rapid succession of collisions prevents the VMS from reaching thermal equilibrium between impacts, potentially causing it to expand beyond its equilibrium radius, which may, in turn, alter its collision cross-section and affect the likelihood and nature of subsequent encounters. The physical and thermodynamical structure of such a VMS lies in a largely unexplored regime, and it may experience significant mass loss through repeated collisions and strong stellar winds, both of which can affect its growth in radius and mass \citep{Dale2006, Suzuki2007}. As a result, our simulation likely provides an optimistic upper limit on the achievable VMS mass under such conditions.

Repeated collisions also rejuvenate the VMS, delaying its natural evolution toward collapse into a BH. VMSs may be important contributors to cosmic hydrogen and helium reionization, a process that depends sensitively on stellar wind mass-loss rates and metallicity \citep{Vink2021, Sander2020}.
In our models, we adopt a relatively high metallicity of $Z = 0.01$, which enhances wind-driven mass loss and makes VMS formation more difficult compared to lower-metallicity environments. Despite this, the VMS forms rapidly enough that winds do not inhibit its growth, owing to the assumption that the star returns to thermal equilibrium following each collision. Between the two codes, the higher collision rate in the \textit{N}-body run leads to more frequent rejuvenation and prolongs the main-sequence lifetime of the VMS compared to the \textsc{MOCCA} run. The stellar wind prescriptions in both simulations are based on the works of \citet{Vink2001}, \citet{Vink2002, Vink2005}, and \citet{Belczynski2010}. As the VMS evolves into the helium-burning phase, it experiences stronger but shorter-lived wind episodes. For this phase, we adopt a luminosity-dependent wind recipe following \citet{Sander2020}. However, since the helium main-sequence lifetime is relatively short ($\sim 0.01~\rm Myr$) for the VMS in these runs, there is no significant mass loss due to winds during this evolutionary stage. In future work, we plan to implement a more refined mass–luminosity relation to improve the modeling of VMS evolution and feedback.

The early and rapid mass gain in our models is essential; had the VMS grown more slowly, stellar winds at $Z = 0.01$ would likely have halted its development, preventing IMBH formation. However, this outcome also depends on how strongly the VMS deviates from thermal equilibrium after collisions, as larger radii could enhance the stellar collision rate and potentially compensate for wind-driven mass loss. Furthermore, the IMBH forms through the direct collapse of the VMS, with an assumed mass loss of approximately 10\% due to neutrino emission \citep{Fryer2012, Kamlah2022a}. However, the exact amount of mass lost, if any, during the transition from an evolved star to a BH remains uncertain.

VMS formation at high mlicities ($Z = 0.01$–0.02) has also been investigated using smoothed-particle hydrodynamics (SPH) simulations with \textsc{ASURA-BRIDGE} \citep{Fujii2024a}. These studies model dense clusters with central densities of $10^6$–$10^9~\rm M_{\odot}\,pc^{-3}$ and show that IMBHs with masses exceeding $4000~\rm M_\odot$ can form within approximately 1.5 Myr \citep{Fujii2024}. The inclusion of gas in these models contributes to the higher central densities and enhances the conditions for runaway collisions and rapid VMS growth. Unlike our gas-free \textit{N}-body and Monte Carlo models, these simulations demonstrate how gas inflows can further support the VMS formation channel. Nevertheless, our results show that even in purely stellar systems, rapid VMS formation is feasible under extreme initial conditions.

If we interpret runaway stellar collisions as analogous to gas accretion, we can compare our results to models of accreting stars \citep{Haemmerle2018, herrington2023}. In particular, we focus on models with an accretion rate of $\sim 0.01~\rm M_{\odot}\,yr^{-1}$, which is comparable to the effective rate at which our VMS gains mass through mergers. These models predict that accreting stars expand to radii of $\sim 10^4~\rm R_\odot$ along the Hayashi track. Notably, \citet{herrington2023} terminate their simulations at central hydrogen depletion. The VMS in our simulations reaches a radius comparable to those found in these accreting-star models. Since stellar radius is an important parameter influencing collision rates and outcomes, the agreement between our VMS radii and those predicted in accretion models supports the plausibility of the VMS growth we observe. However, the internal structure of a VMS built through collisions may differ significantly from that of an accreting MS star. In particular, collisional VMSs may develop a dense core with an extended, dilute envelope, potentially reducing the efficiency of further mass buildup by lowering the effective collision cross-section. This structural difference adds uncertainty when interpreting the final VMS mass in our simulations.

Furthermore, \citet{Martinet2023} present stellar evolution models computed with the \textsc{GENEC} code \citep{Eggenberger2008} for metallicity $Z = 0.014$, including both rotating and non-rotating stars with initial masses of 180, 250, and $300~\rm M_\odot$. These models yield main-sequence lifetimes of approximately 2.2–2.6 Myr. In comparison, the \textsc{SSE} models by \citet{Hurley2000}, which we use in our simulations, give a main-sequence lifetime of 2.9 Myr at $Z = 0.01$. This agreement suggests that the stellar evolution prescriptions adopted in our simulations are consistent with more detailed stellar evolution models, lending credibility to our estimates of VMS lifetimes and their impact on subsequent cluster evolution.

The radiative efficiency of stellar winds from VMSs could play an important role in the chemical enrichment of star clusters. This process has been proposed as a potential mechanism for the formation of multiple stellar populations (MSPs) \citep{Gieles2018, Gieles2025}. Recent high-resolution hydrodynamic simulations have provided the first detailed insights into VMS self-enrichment in massive clusters, highlighting its relevance for MSP formation. Stellar winds enriched with elements such as Na, Al, and N can be retained within the cluster and contribute to the formation of chemically distinct stars within just a few million years \citep{Lahe2024}. In addition to stellar winds, stellar mergers offer an alternative pathway for MSP formation.
In dense clusters where mergers are more frequent, these events can eject chemically enriched material into the surrounding gas. This material—whether expelled through dynamical interactions, explosions, or fusion-driven outflows—can lead to the formation of new stars with distinct chemical compositions \citep{Wang2020e}. Recent observational studies have also revealed detailed kinematic differences between stellar populations, showing that second-generation stars tend to rotate more rapidly and are more centrally concentrated than first-generation stars \citep{Dalessandro2024}.
These findings provide further motivation for detailed modeling of early cluster enrichment scenarios involving VMSs, especially in young massive clusters forming at high redshift.

\subsection{Observational context for VMS formation at high redshift}\label{sec:summary-observations}

Observing massive, compact, and low-metallicity star clusters around the time of core collapse remains extremely challenging. However, high-redshift galaxies are widely expected to form such clusters efficiently due to their high gas fractions and low mlicities \citep[e.g.][]{Tacconi2010}, as well as their clumpy substructures that favor gravitational fragmentation and rapid star formation \citep[e.g.][]{Devereaux2024}. These conditions closely mirror those adopted in our simulations, reinforcing the plausibility of early VMS formation via runaway stellar collisions. We note that the majority of massive stars observed have masses of approximately $100-300~\rm M_\odot$ \citep{Doran2013, Keszthelyi2025}, stars with masses above this limit remain purely theoretical.

Recent observations using strong gravitational lensing have begun to probe star-forming clumps and stellar clusters in high-redshift galaxies. For example, \citet{Claeyssens2023} analyzed clump populations in galaxies at $z = 1$–$8.5$, identifying half-mass radii ranging from a few to several tens of parsecs. They also found evidence for a mass–size scaling relation that may lie above the local relation \citep{Brown2021}, although the observed population is likely biased toward more massive and extended systems. Nevertheless, extremely compact clusters are still expected to form in high-density environments, even if they remain observationally elusive.

From a theoretical standpoint, the Toomre instability criterion implies that the typical cluster size should scale inversely with the square root of the gas surface density, i.e., $R_h \propto \Sigma_{\mathrm{gas}}^{-1/2}$ \citep{Toomre1964}. This suggests that star clusters formed in gas-rich, high-redshift galaxies should be significantly more compact than their local counterparts—a trend also supported by simulations and nearby observations \citep{Choksi2021}. Reinforcing this, \citet{Adamo2024} recently reported several compact clusters at $z \sim 10$ with effective radii below one parsec and enhanced surface densities compared to local analogues. These findings support the idea that extremely dense clusters, such as those used in our models, could plausibly form in the early Universe. However, our models do not include residual gas, which could remain trapped in such compact systems and influence early dynamical evolution and VMS formation.

The presence of such dense clusters may also be required to explain chemical signatures observed in some high-redshift systems. Notably, the candidate galaxy GN-z11 at $z = 10.6$ \citep{Bouwens2010, Tacchella2023, Nagele2023, Maiolino2024} exhibits an unusually high nitrogen-to-oxygen abundance ratio, with $\log_{10} (\mathrm{N}/\mathrm{O}) > -0.25$—a value roughly four times the solar ratio \citep{Cameron2023}. Similar enhancements have been reported in other high-$z$ sources \citep{Marques-Chaves2024}.
Rapidly rotating Pop-III stars have been suggested as a potential source of nitrogen enrichment \citep{Tsiatsiou2024, Nandal2024}. Additionally, high nitrogen abundances could originate from hydrogen-burning nucleosynthesis in VMS formed via runaway stellar collisions \citep{Charbonnel2023}. Our results, which show VMS growth within a few Myr support this scenario by demonstrating that VMSs can form early and potentially enrich their surroundings before core collapse. 

Compact massive clusters may also form during major galaxy mergers, particularly at high redshift where galaxies are more gas-rich and m-poor. Mergers can trigger large-scale inflows of gas, leading to starbursts and the formation of dense stellar systems containing $>10^4$ to $10^6$ stars \citep{Renaud2015}. While nearby merging systems like the Antennae galaxies have near-solar mlicities \citep{Fall2005}, similar processes occurring in early galaxies \citep[e.g.][]{Pettini1997} could create the ideal conditions for VMS formation. This has been demonstrated in simulations by \citet{Lahe2019, Lahe2020a, Lahe2020}, where massive clusters with masses $\gtrsim 10^5~\rm M_\odot$ formed in the aftermath of gas-rich dwarf galaxy mergers. These simulated clusters are broadly consistent with the initial conditions explored in our models, reinforcing the relevance of our results for understanding early IMBH seed formation in the cosmological context.

\subsection{Outlook: Observational Signatures and Future Tests}\label{sec:summary-outlook}

Tidal disruption events (TDEs) represent a promising observational signature of massive black holes (BHs), including those that may have formed via the collisional pathways modeled in this work. These events occur when a star passes sufficiently close to a BH to be partially or fully disrupted and accreted, producing luminous flares across the electromagnetic spectrum \citep{Hils1975, Rees1988}. TDEs are associated with the so-called “loss cone”—a region of phase space containing stellar orbits that bring stars close enough to be disrupted \citep{Chandrasekhar1942}. In spherical systems, this region is rapidly depleted, and its replenishment through two-body relaxation is typically slow \citep{Broggi2024}. Although TDEs can contribute to BH growth, our simulations conclude shortly after IMBH formation, and longer-term dynamical evolution—including loss cone refilling—remains a key area for future study.

In the “empty loss cone” regime, where stellar orbits rarely intersect the BH’s tidal radius, BH growth stalls unless aided by additional processes such as gas accretion or compact object mergers \citep[e.g.][]{Stone2016, Ryu2020a, Ryu2020b, Ryu2020, Stone2020}. One particularly intriguing channel involves the gradual inspiral of compact remnants (neutron stars or stellar-mass BHs) around an IMBH—a class of events known as extreme mass ratio inspirals (EMRIs) \citep{Hils1995}. EMRIs are shaped by dynamical friction, relativistic precession, and tidal forces, and they represent a key gravitational wave (GW) signature of IMBHs in dense star clusters \citep{Broggi2022}. Our results, which demonstrate early IMBH formation via runaway collisions, suggest that such GW sources could appear much earlier in cosmic history than traditionally assumed.

Crucially, these processes are not only theoretically predicted—they are increasingly accessible to observation. TDEs produce luminous flares observable with facilities such as the Hubble Space Telescope (\href{https://hubblesite.org/home}{HST}) \citep{Leloudas2016} and the Neil Gehrels Swift Observatory (\href{https://swift.gsfc.nasa.gov/}{Swift}) \citep{Brown2016}. Meanwhile, mergers involving compact objects—including EMRIs—are detectable with current GW observatories such as \href{https://www.ligo.caltech.edu/page/ligos-ifo}{LIGO}, \href{https://www.virgo-gw.eu/}{Virgo}, and \href{https://gwcenter.icrr.u-tokyo.ac.jp/en/}{KAGRA} \citep{Abbott2024}, and will be prime targets for future space-based missions like \href{https://lisa.nasa.gov}{LISA} \citep{Amaro-Seoane2017, McCaffrey2025} and the \href{https://www.et-gw.eu/}{Einstein Telescope (ET)} \citep{Punturo2010}.

Complementary electromagnetic observations will further aid in testing the predictions of our collisional IMBH formation scenario. The upcoming \href{http://jasmine.nao.ac.jp/index-en.html}{JASMINE} mission will offer ultra-precise astrometry of the Milky Way’s nuclear star cluster, enabling dynamical BH mass measurements. The \href{https://elt.eso.org}{Extremely Large Telescope (ELT)}, with its near-infrared imager \href{https://elt.eso.org/instrument/MICADO/}{MICADO}, will be capable of resolving the sphere of influence of BHs at distances up to twice that of JWST \citep{Davies2018}. Already, JWST has revealed a population of compact, red sources—the so-called "Little Red Dots"—which are interpreted as dusty supermassive BHs (SMBHs) or obscured, star-forming galaxies \citep{Kokorev2024, Akins2025, Matthee2024, Napolitano2025}. These high-redshift objects exhibit unusually large BH-to-stellar mass ratios, up to two orders of magnitude higher than those seen locally \citep{Goulding2023, Scoggins2023, Ubler2024, Furtak2024, Juodzbalis2024}. Such findings suggest the presence of "overmassive" BHs in the early Universe and challenge conventional models of BH seed formation.

Our results offer a complementary explanation: the formation of massive BH seeds via runaway stellar collisions in dense clusters, followed by early collapse into IMBHs. This purely collisional pathway may help reconcile observations of overmassive BHs at high redshift with theoretical models, particularly if dense star clusters like those in our simulations were common in the early Universe. Future multi-messenger observations—combining GWs and electromagnetic signatures—will be essential for distinguishing between different IMBH formation channels and testing the predictions of this work.

%

\begin{acknowledgements}

MCV acknowledges funding through ANID (Doctorado acuerdo bilateral DAAD/62210038) and DAAD (funding program number 57600326). Furthermore, acknowledges to the International Max Planck Research School for Astronomy and Cosmic Physics at the University of Heidelberg (IMPRS-HD). 

AA acknowledges support for this paper from project No. 2021/43/P/ST9/03167 co-funded by the Polish National Science Center (NCN) and the European Union Framework Programme for Research and Innovation Horizon 2020 under the Marie Skłodowska-Curie grant agreement No. 945339. For the purpose of Open Access, the authors have applied for a CC-BY public copyright license to any Author Accepted Manuscript (AAM) version arising from this submission.

AWHK and RS acknowledge NAOC International Cooperation Office for its support in 2023, 2024, and 2025. RS acknowledges the support by the National Natural Science Foundation of China (NSFC) under grant No. 12473017. This research was supported in part by grant NSF PHY-2309135 to the Kavli Institute for Theoretical Physics (KITP).

NH is a fellow of the International Max Planck Research School for Astronomy and Cosmic Physics at the University of Heidelberg (IMPRS-HD). NH received financial support from the European Union's HORIZON-MSCA-2021-SE-01 Research and Innovation programme under the Marie Sklodowska-Curie grant agreement number 101086388 - Project acronym: LACEGAL. 

This material is based upon work supported by Tamkeen under the NYU Abu Dhabi Research Institute grant CASS. FFD and RS acknowledge support by the German Science Foundation (DFG, project Sp 345/24-1). 

PB thanks the support from the special program of the Polish Academy of Sciences and the U.S. National Academy of Sciences under the Long-term program to support Ukrainian research teams grant No.~ PAN.BFB.S.BWZ.329.022.2023. The work of PB was also supported by the grant No.~BR24992759 ``Development of the concept for the first Kazakhstan orbital cislunar telescope - Phase I'', financed by the Ministry of Science and Higher Education of the Republic of Kazakhstan.

MAS acknowledges funding from the European Union’s Horizon 2020 research and innovation programme under the Marie Skłodowska-Curie grant agreement No.~101025436 (project GRACE-BH, PI: Manuel Arca Sedda). MAS acknowledge financial support from the MERAC foundation.

AT is supported by Grants-in-Aid for Scientific Research (grant No. 19K03907 and 24K07040) from the Japan Society for the Promotion of Science.

DRGS gratefully acknowledges support by the ANID BASAL project FB21003, as well as via Fondecyt Regular (project code 1201280) and ANID QUIMAL220002. DRGS thanks for funding via the  Alexander von Humboldt - Foundation, Bonn, Germany.

AH and MG were supported by the Polish National Science Center (NCN) through the grant 2021/41/B/ST9/01191.

Xiaoying Pang acknowledges the financial support of the National Natural Science Foundation of China through grants 12173029 and 12233013.

TN acknowledges the support of the Deutsche Forschungsgemeinschaft (DFG, German Research Foundation) under Germany’s Excellence Strategy - EXC-2094 - 390783311 of the DFG Cluster of Excellence ''ORIGINS''.

RC is supported in part by the National Key Research and Development Program of China and the Zhejiang provincial top level research support program.

AE acknowledge financial support from  the Center for Astrophysics and Associated Technologies CATA (FB210003).

Computations were performed on the HPC system Raven at the Max Planck Computing and Data Facility, and we also acknowledge the Gauss Centre for Supercomputing e.V. for computing time through the John von Neumann Institute for Computing (NIC) on the GCS Supercomputer JUWELS Booster at Jülich Supercomputing Centre (JSC).

Finally, we also acknowledge A. Sander and his team for helpful comments.

\end{acknowledgements}


\section*{Data Availability}
The underlying data, including the initial model used in this work, as well as the output and diagnostic files from both the \textsc{Nbody6++GPU} and \textsc{MOCCA} simulations, are publicly available at \href{https://doi.org/10.5281/zenodo.15283075}{https://doi.org/10.5281/zenodo.15283075}.
The \href{https://github.com/kaiwu-astro/Nbody6PPGPU-beijing}{\textsc{Nbody6++GPU}} code version that includes the \texttt{level C} stellar evolution prescriptions \citep{Kamlah2022a} is publicly available.
The \href{https://github.com/agostinolev/mcluster.git}{\textsc{McLuster}} version used to generate the initial conditions is also publicly available as described in \citet{Leveque2022a, leveque2022}.


\bibliographystyle{aa} 
\bibliography{ref2}


\begin{appendix} 

\section{Stellar evolution treatment}
\label{sec:evolution_parameters}

In our work, we follow the \texttt{level C} stellar evolution package, as presented in \citep[see Table X in][]{Kamlah2022a} and the stellar evolution fitting formulae from \textsc{SSE} by \citet{Hurley2000}, which also describes the stellar evolution routines and parameters in detail, with updates as described in \citet{Spurzem2023} and in Sections below.

We use the metallicity dependent winds following \citet{Vink2001,Vink2002,Vink2005,Belczynski2010} across the full mass range. 

Here, we use the recipe for the stellar winds that does not take into account the bi-stability jump \citep{Belczynski2010}, and explained in the following. We highlight here that the winds of the massive (MS) stars have enormous effect on our results. The mass growth of the massive MS star via collisions is in constant competition with the mass loss driven by the strong, hot, fast OB-type winds for massive MS stars \citep{Vink2021}. 
We have a strong mass loss recipe for massive OB-type MS stars which takes into account the so-called bi-stability. The bi-stability jump causes a sudden jump in the line-driving characteristic of Iron. Specifically, this is the result of a recombination of the Fe(IV) to the Fe(III) ion at effective stellar temperatures of around $T\sim 25\,000$~K. The Fe(III) ion is a much more effective driver for radiative transport and therefore, the mass loss is enhanced \citep[for more details see e.g.][]{Belczynski2010,Vink2021,Bjorklund2023}. This mass loss recipe can be chosen in the future, due to accumulating evidence that the bi-stability jump actually exists in nature \citep[see e.g. recent observations by][and sources therein]{Bernini-Peron2023}, even though the exact impact is still unexplored in star cluster dynamics.\\

For the compact objects evolution, we use remnant mass prescriptions following \cite{Fryer2012}, and here we choose the delayed SNe mechanism as the slow extreme of the convection-enhanced, neutrino-driven, SNe paradigm. We use standard momentum conserving fallback-scaled kicks (drawn from a Maxwellian distribution with a dispersion of 265.0~$\mathrm{km\,s}^{-1}$ from \citet{Hobbs2005}) for the NSs and BHs \citep{Belczynski2008}. This is not the case for the NSs and BHs produced by the electron-capture SNe (ECSNe), accretion-induced collapse (AIC) and merger-induced collapse (MIC) \citep{Podsiadlowski2004,Ivanova2008,Gessner2018,Leung2020} and that are subject to low velocity kicks (drawn from a Maxwellian distribution with a dispersion of 3.0~$\mathrm{km\,s}^{-1}$ from \citet{Gessner2018}). \\

For the natal spins of the compact objects, we have(i) BHs receive natal spins following the \texttt{Geneva} models \citep{Banerjee2020,Banerjee2021} and (ii) WDs receive natal kicks following \citet{Fellhauer2003} (drawn from a Maxwellian distribution with a dispersion of 2.0~$\mathrm{km\,s}^{-1}$ but, which is capped at 6.0~$\mathrm{km\,s}^{-1}$). We switch on the weak (pulsational) pair instability SNe ((P)PISNe) following \citet{Leung2019}.\\
Finally (iii) the NS did not receive any upgrade, following \citep{Hurley2000}, where the spin is derived from the classical polytrope equation for n=3/2.
Lastly, we briefly outline the settings concerning the runs with general relativity merger recoil kicks. When there is a collision between two compact objects (in these simulations there will be NSs or BHs), then the merger product will have a spin drawn from a Maxwellian distribution with $\sigma=0.2$. In the collision itself, there is a mass loss due to the emission of GWs. Therefore, the final merger product will have 0.985 times the sum of the masses of the compact objects that participated in the merger. The kick velocity is then calculated as described in \citet{ArcaSedda2023}. We stress that these kicks apply to all compact object mergers, so also compact object binary mergers consisting of WDs and NSs.

For direct collisions with mergers between two stars, which are not remnants, we assume that the new star assumes its new equilibrium instantaneously afterwards, which is one of the important simplifications to be discussed further in Sect.~\ref{sec:summary}.

\subsection{Treatment of stellar winds for He star}
\label{sec:stellar_winds_He_star}
The formation of VMS and posterior collapse as BH, is highly influenced by stellar winds, when it becomes an He star, which vary depending on the star metallicity. We updated the old treatment used in \textit{N}-body for stellar wind from \citet{Hamann1998, Vink2005} with the recipe from \citet{Sander2020},

\begin{equation}
\dot{M} = \dot{M}_{10} \left( \log \frac{L_{Edd}}{L_0} \right)^{\alpha} \left( \frac{L_{Edd}}{10L_0} \right)^{3/4}
\end{equation} 
Where

\begin{align}
L_0 &= 10^{5.06} \left(Z/Z_\odot\right)^{-0.87}, \\
\alpha  &= 0.32 \log\left(Z/Z_\odot\right) + 1.4, \\
\dot{M}_{10} &= 10^{-4.06} \left(Z/Z_\odot\right)^{-0.75},
\end{align} and $L_{Edd}$ is the Eddington luminosity.

\section{Upgrade of stellar radius evolution}
\label{sec:radius_evolution}

Stellar evolution parameters, including the stellar radius, are provided by the \textsc{SSE} algorithm \citep{Hurley2000} with updates as described in \citet{Kamlah2022a}. 
This algorithm was fitted to the results of detailed stellar models of up to 50.0~$\mathrm{M}_{\odot}$ and tested for reliable behaviour up to 100.0~$\mathrm{M}_{\odot}$. 
Thus, using \textsc{SSE} beyond this mass range was historically at the peril of the user and inconsistencies in the stellar radius for masses well above 100.0~$\mathrm{M}_{\odot}$ had been noted previously (e.g. Gieles, private correspondence). 
For the purpose of the \textsc{Nbody6++GPU} models reported in this work, we therefore conducted a series of tests of \textsc{SSE} in the VMS regime and implemented some minor additions to \textsc{SSE} to safeguard the routines when used for VMS evolution. In Appendix~\ref{sec:appsevcomp}, including Fig.~\ref{fig:radius}, we explain in some more detail how the upgrade has been done. Note that this is aimed primarily at making the routines numerically safe to use as well as removing any obvious inconsistencies. In the long term, we plan to improve the code with routines for very low mlicities \citep{Tanikawa2020} and massive star formation.

\section{Upgrade of rejuvenation via stellar collisions}
    \label{sec:rejuvenation_collisions}

    An important consequence of stellar evolution of a massive star is that the intermediate stages of stellar evolution have a short lifetime before the star enters in the final phases of its life. From 20 $\mathrm{M}_{\odot}{}$ (although a precise threshold limit has still not agreed by the scientific community), a star can evolve into a BH, and larger masses would evolve into BHs even faster. VMS stars, then, would have an even shorter timescale, but this process can be slowed down through a rejuvenation mechanism.\\  
    We use the \textsc{BSE} package by \citet{Hurley2002a,Hurley2005}, with upgrades related to the treatment of rejuvenation of main sequence stars upon collision with other MS stars and mass loss during hyperbolic collisions, as will be described in this Appendix. Rejuvenation originally was intended to describe the formation of blue straggler stars after stellar collisions \citep{Hurley2005}. These stars lie above a stellar populations main sequence turn-off point \citep[see e.g., the observational evidence from][]{Giesers2018}. This observation implies that the effective age, $A_{\mathrm{MS}}$, of these MS stars after a collision and a merger, is younger than the average age of the stellar population. This concept (described in more detail below) delivers unphysical results for collisions of a VMS with small mass MS stars, i.e. those where the mass ratio $q$,
    \begin{equation}
    \label{eq:massratio}
    q = \frac{M_{\mathrm{MS,2}}}{M_{\mathrm{MS,1}}} \quad ,
    \end{equation}
    becomes very small ($q\ll 1$). In the following, we will analyze in more detail the traditional rejuvenation treatment 
    \citep[Eq.~\ref{eq:age3}, see also discussion in][]{Hurley2005}, explain why it fails for very small $q$, and describe our workaround (Eq.~\ref{eq:age3_correction}). 

    In the context of an active mass transfer from one MS star to another MS star, the primary (and more massive) MS star will be rejuvenated upon reception of mass transfer (assuming the instantaneous mixing of the stellar material). The degree of mixing strongly depends on the mass of the star, which changes from fully convective smaller masses to only convective cores for larger masses:

    \begin{enumerate}
        \item MS stars with convective cores $\left(M>1.25~\mathrm{M}_{\odot}\right)$: upon collision and subsequent mass gain by the primary MS star, its core grows in size. The latter leads to enhanced convective mixing of unburnt hydrogen fuel from the outer stellar layers to the core. This process will lead to the rejuvenation of the star. \textsc{BSE} numerically translate this rejuvenation by conserving the amount of burnt hydrogen and grown the stellar core of the primary star. The new age value is thus directly proportionally to the remaining fraction of unburnt hydrogen at the centre.
        \item MS stars with radiative cores $\left(0.35 \leqslant M / \mathrm{M}_{\odot} \leqslant 1.25\right)$: the hydrogen burning process in the core is affected slightly, which leads to a decrease in the effective age of the star upon collision. In contrast to high-mass MS stars (with convective cores and very low-mass fully convective stars), the age of the star is altered so that the fraction of the MS life-time, which has elapsed, is unchanged by the change of mass due to the inefficient mixing.
        \item fully convective MS stars $\left(M<0.35~ \mathrm{M}_{\odot}\right)$: these low-mass stars are treated in the same way as the stars with convective cores.
    \end{enumerate}

If we work under the base assumption that hydrogen in the MS core is uniformly distributed after a collision, which in turn is a result of the assumption of instantaneous (convective) mixing of the hydrogen fuel from envelope to core, the \textsc{BSE} code terminates the MS phase of the star and enters the Hertzsprung gap (HG) phase after 10~\% of the total amount of hydrogen in the MS star have been burnt \citep{Hurley2000,Hurley2005}. \\

The computational application is described in the following.
We first consider the collision of two MS stars with (effective) ages $A_{\mathrm{MS,1}}$ and $A_{\mathrm{MS,2}}$, MS life-times $\tau_{\mathrm{MS,1}}$ and $\tau_{\mathrm{MS,2}}$, as well as masses $M_{\mathrm{MS,1}}$ and $M_{\mathrm{MS,2}}$, respectively. Here, the subscripts $1$ and $2$ denote the primary (and thus more massive) MS star and the secondary, respectively. The collision produces the "third" MS star with effective age $A_{\mathrm{MS,3}}$, MS life-time $\tau_{\mathrm{MS,3}}$ and mass $M_{\mathrm{MS,3}}$. If we know these parameters, \citep[like suggested in][]{Hurley2005} we can evaluate the final effective age $A_{\mathrm{H,MS,3}}$ of the produced MS star with the following equation \citep[see also][for more information]{Glebbeek2008}:

\begin{equation}
    \label{eq:age3}
    A_{\mathrm{H,MS,3}} = \frac{0.1}{1\!+\!q}\cdot \tau_{\mathrm{MS,3}} \cdot
    \left(  \frac{A_{\mathrm{MS,1}}}{\tau_{\mathrm{MS,1}}} + q \cdot 
            \frac{A_{\mathrm{MS,2}}}{\tau_{\mathrm{MS,2}}} \right),
\end{equation}

where the factor of 0.1 takes into account the previously mentioned condition that the MS phase ends after 10~\% of the total amount of hydrogen in the MS star have been burnt. We have found that Eq.~\ref{eq:age3} works well for MS collisions that have mass ratios $q \approx 1$, which might be expected for many blue stragglers. \\ 

However, for the extreme case $q\rightarrow 0$ (i.e., VMS primary star and MS secondary star) we get, from the traditional formula,
\begin{equation}
    A_{\mathrm{H,MS,3}} \approx 0.1 \cdot A_{\mathrm{MS,1}}  \ ,
\end{equation}
since $\tau_{\mathrm{MS,3}} \approx \tau_{\mathrm{MS,1}} $. So such collisions would lead to progressive strong nonphysical rejuvenation of a VMS. A sequence of such collisions could effectively prevent the transformation of a VMS to a BH for a large timescale, especially in very dense star clusters.
As a result, if we have the growth of a massive MS star with one or more collisions of near zero $q$, the massive MS star would never age and never reach the Hertzsprung gap as long as the frequency of collisions stays constant as the rejuvenation of the star would be much too large. This would mean that star clusters with larger core densities unnaturally form (young) massive stars which stay in the MS and an almost constant radius.

Such nonphysical age reduction in the previous rejuvenation process has another unwelcome consequence, where the stellar radius of the star is always reset to a value that corresponds to the star's effective age. However, a MS star grows in size as it ages and approaches the Hertzsprung gap \citep[see e.g.][]{Pols1998,Hurley2000,Tanikawa2020}. The stellar size increment accelerates the closer the star is located to the Hertzsprung gap. Therefore, the star will increase in size measured by the stellar radius in several orders of magnitude for VMSs. This implies that a MS star produced via repeated stellar collisions, whose effective age is treated with Eq.~\ref{eq:age3}, will be much too small, which in turn greatly also affects its scattering cross-section and therefore, its mass growth, in dense stellar environment. \\

In our previous direct $N$-body works, where only a limited number of stellar collisions have been produced, no MS star merger chains have been produced that go beyond 10 compared with the densest star cluster model presented here with $r_{\mathrm{h,i}}=0.1~(\mathrm{pc})$  collisions \citep[see e.g.][for recent direct $N$-body work, which includes comparable astrophysical complexity as the work presented here, i.e. full stellar evolution of single and binary stars, tidal fields, mass spectrum and so on.]{Rizzuto2021,Rizzuto2022,Kamlah2022a,Kamlah2022b,ArcaSedda2023,ArcaSedda2024,ArcaSedda2024a}. 
Therefore the aforementioned rejuvenation issue was not a problem in these papers.
However, for this paper we have hundreds of such mergers with small $q$, therefore a simple workaround had to be applied for the traditional treatment. We now use a new expression $A_{\mathrm{MS,3}}$ for the age of the merger product as:

\begin{equation}
    \label{eq:age3_correction}
    A_{\mathrm{MS,3}} = A_{\mathrm{MS,1}} - q \cdot (A_{\mathrm{MS,1}} - A_{\mathrm{H,MS,3}}).
\end{equation}
This treatment for the age of the merged body ensures that (i) if $q\rightarrow 1$ (i.e., the two colliding stars have similar masses) we get Eq.~\ref{eq:age3} (i.e. the age of the merger product is taken from the classical formula) , while (ii) for $q\rightarrow 0$ (i.e., the primary star is much more massive than the secondary) the age of the VMS remains nearly unchanged. Note that \cite{Mapelli2016} used a different rejuvenation algorithm based on earlier papers than \cite{Hurley2005,Glebbeek2008}, which is less general than our algorithms.

Unlike the original treatment in the direct $N$-body simulations using \textsc{Nbody4}, as presented in \citet{Hurley2005}, we incorporate in  mass loss for hyperbolic collisions, so collisions where the sum of the binding energy of the binary and orbital kinetic energy are positive. If we have a hyperbolic collision between two MS stars, then the code reduces the gained mass by the primary depending on the total energy in the two body center of mass and the binding energy of the secondary, resulting in: 
\begin{equation}
   \delta M_{\mathrm{coll}}  = M_{\mathrm{MS,2}} \cdot \frac{1}{1+|E_2 / E_{\rm tot}|}
\end{equation}
where $\delta M_{\mathrm{coll}}$ is the mass loss of the secondary star due to the collision of the secondary star; $E_{\rm tot}$ is the total energy in the two-body encounter and $E_2$ the binding energy of the secondary star. The remaining mass of $M_{\mathrm{MS,2}}$ is gained by the primary star. No mass loss of the primary star is taken into account currently (we are assuming here $q\ll 1$). Note that in all previously published versions of \textsc{Nbody6++GPU} and earlier versions of \textsc{Nbody6} a very much simplified mass loss for hyperbolic collisions was assumed - just a constant mass loss of 30\% of the secondary mass for hyperbolic collisions.

\section{Comparison of stellar radii old and upgraded}
\label{sec:appsevcomp}

In \textsc{SSE} the radius of a star at the end of the main sequence is given by the function $R_{\rm TMS}$ \citep[Eqs. 9a and 9b]{Hurley2000} which can reach very high values in the VMS range: $2\!\cdot\!10^7 \, \mathrm{R}_{\odot}$ for 50,000~$\mathrm{M}_{\odot}$ at $Z = 0.01$. 
We added an extra function to flatten this out for high masses ($M > 3,000 \, \mathrm{M}_{\odot}$) where: 
\begin{equation}
    R_1 = 6\cdot10^4 + 2.8 \frac{M}{\mathrm{M}_{\odot}}\, \mathrm{R}_{\odot}, \quad
    R_2 = 2\cdot10^5 \, \mathrm{R}_{\odot}
\end{equation}

\begin{equation}
    R_{\rm TMS} = \min \left( R_{\rm TMS}, R_1, R_2 \right)\, \mathrm{R}_{\odot}\label{eq:RTMS}
\end{equation}

\begin{figure*}[h!]
\centering
\begin{subfigure}[t]{0.49\textwidth}
    \centering
    \includegraphics[width=\textwidth]{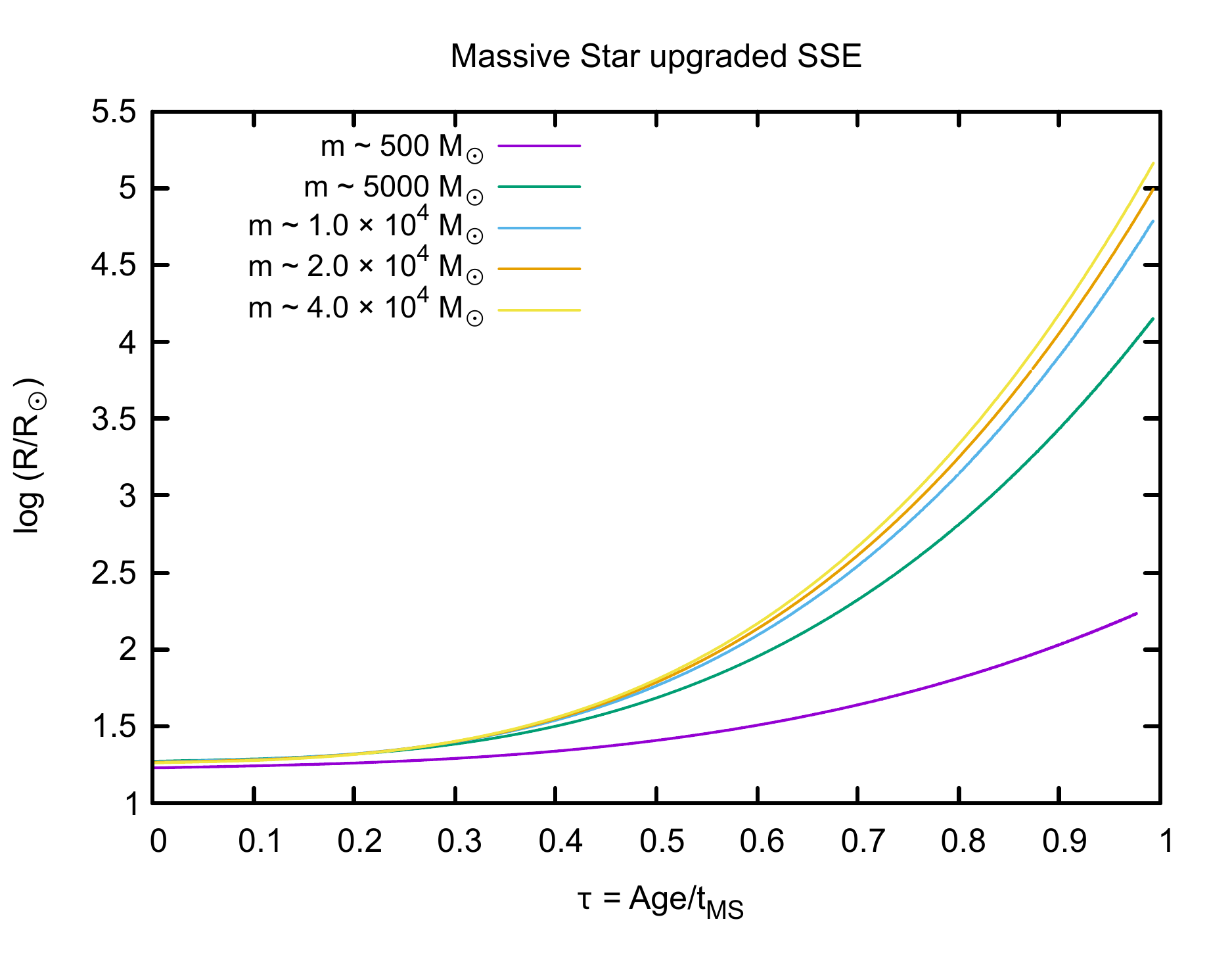}
\end{subfigure}
\hfill
\begin{subfigure}[t]{0.49\textwidth}
    \centering
    \includegraphics[width=\textwidth]{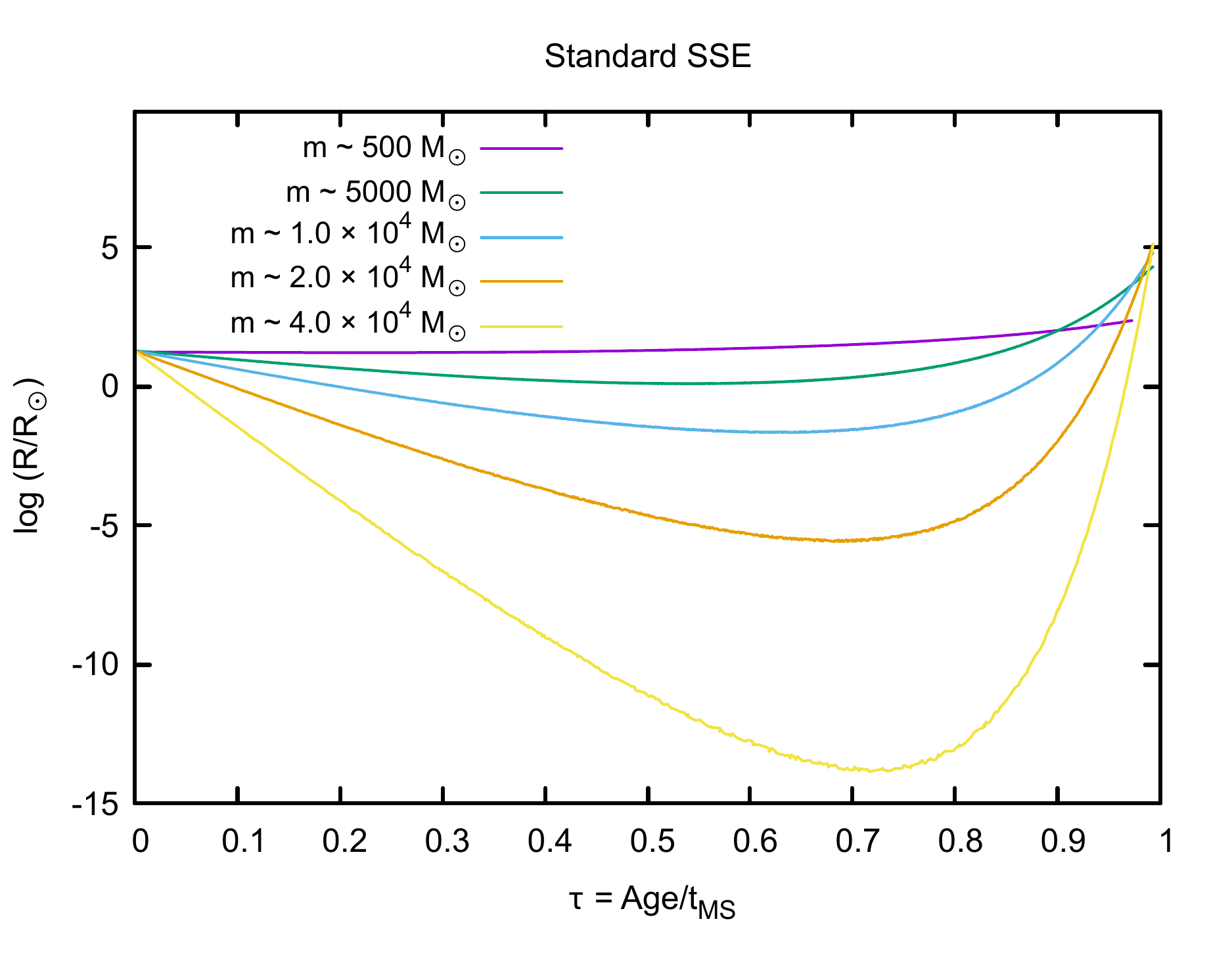}
\end{subfigure}
\caption{Radii of massive stars (five masses as given in key) as a function of age in units of main sequence lifetime; right panel: results by standard SSE; left panel: results from our upgraded SSE.}
\label{fig:radius}
\end{figure*}

This ensures that $R_{\rm TMS}$ cannot become larger than a maximum of $2 \cdot 10^5 \, \mathrm{R}_{\odot}$ which is reached for 50,000~$\mathrm{M}_{\odot}$. This roughly coincides with the radii of VMS models with $M/{\rm M}_{\odot} \ge 10^4$, at in the beginning of the Kelvin-Helmholtz contraction phase \citep{Fricke1973,Hara1978}. 
(cf. Fig.~\ref{fig:radius}).
It was also necessary to impose that the coefficient $a_{24}$ in Eq.~9b of \citet{Hurley2000} could not become negative (which could previously occur at low $Z$ and cause problems with the radius calculation for VMSs). The term $a_{24}$ is an interpolation term for the $R_{\rm TMS}$. 
The function to describe the evolution of the radius on the MS is quite complicated \citep[Eq. 13 and associated equations]{Hurley2000}, involving a number of higher order terms and a component aimed at producing the MS hook feature. 
All of which has the potential to create inconsistencies for very high masses so we simplified the treatment for $M > 100 \, \mathrm{M}_{\odot}$ by phasing out the hook component (allowing $\Delta R$ to go smoothly to zero across the range $100 < M/\mathrm{M}_{\odot} < 120$), allowing the $\beta$ term to go to zero (across the same mass range) and capping the $\alpha$ term at its $100 \, \mathrm{M}_{\odot}$ value. 
The $\gamma$ term in Eq. 13 of \citet{Hurley2000} is already zero in this high mass range. Fig.~\ref{fig:radius} illustrates the improvements for $M/{\rm M}_{\odot} > 500$, showing that we now get a steady increase of stellar radius with time up to the limit defined by Eq.~\ref{eq:RTMS}. Note that the left panel in Fig.~\ref{fig:radius} is solely for illustrative purposes, to show what happens if standard SSE would be used for such high masses, for which it has been never intended.

\section{Comparison of ageing formulae}

\begin{figure}[hb!]
  \centering
  \includegraphics[width=\columnwidth]{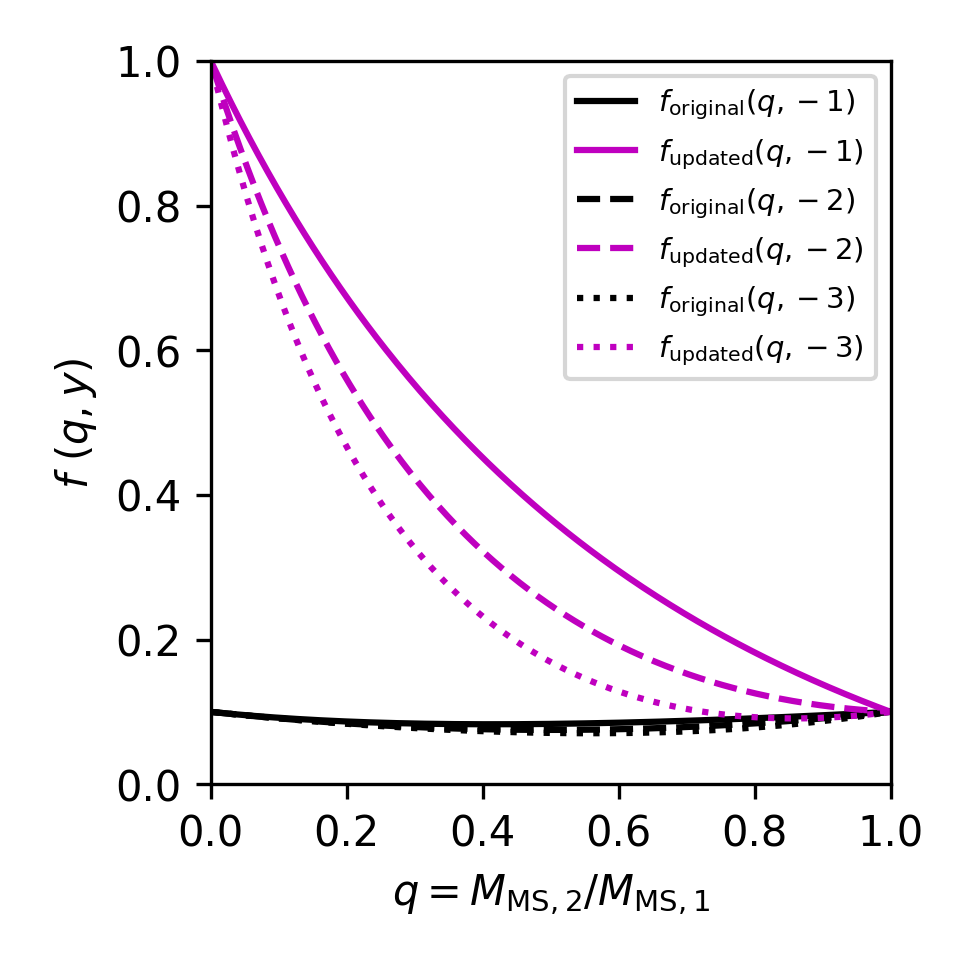}
  \caption{%
    Figure showing the family of functions for the original treatment $f_{\mathrm{original}}(q,y)$  and the family of functions with our updated treatment $f_{\mathrm{updated}}(q,y)$ for three distinct values of $y \in (-1,-2,-3)$ against the mass ratio $q$ of the MS star collision parnters.
  }
  \label{fig:function_family_age3.jpg}
\end{figure}

Already for $q<0.75$, Eq.~\ref{eq:age3} gives too much rejuvenation to the resulting blue straggler, as can be seen in Fig.~\ref{fig:function_family_age3.jpg}. The details of Fig.~\ref{fig:function_family_age3.jpg} are explained in the remainder of this section.

A direct comparison between the original treatment in Eq.~\ref{eq:age3} and our updated treatment Eq.~\ref{eq:age3_correction} is shown in Fig.~\ref{fig:function_family_age3.jpg}. Here the dependence of a function $f(q,y)$ on the mass ratio $q$ from Eq.~\ref{eq:massratio} is shown for three separate values of $y$. We arrive at the functional of $f(q,y)$ as follows:
firstly, we assume that there is no mass in collisions, which implies that 
\begin{equation}
    \frac{1}{1\!+\!q} = \frac{M_{\mathrm{MS,1}}}{M_{\mathrm{MS,3}}}.
\end{equation}
Next, we define two quantities to achieve a more compact functional form:
\begin{equation}
    \label{eq:X}
    X = \left( \frac{A_{\mathrm{MS,1}}}{\tau_{\mathrm{MS,1}}} + q \cdot \frac{A_{\mathrm{MS,2}}}{\tau_{\mathrm{MS,2}}} \right)  \ \ ; \ \ 
    Y = \frac{A_{\mathrm{MS,3}}}{\tau_{\mathrm{MS,3}}}
\end{equation}
Using this notation, Eq.~\ref{eq:age3} becomes
\begin{equation}
    \label{eq:Yoriginal}
    Y_{\mathrm{original}} = 0.1 \cdot \frac{1}{1\!+\!q} \cdot X .
\end{equation}
Note that \cite{Mapelli2016} use a rejuvenation algorithm based on earlier papers, in which $X = A_{\mathrm{MS,1}}/\tau_{\mathrm{MS,1}}$ only (no dependence on secondary parameters), and a factor 1.0 instead of 0.1 in Eq.~\ref{eq:Yoriginal} was used.  
Our new model of rejuvenation Eq.~\ref{eq:age3_correction} becomes
\begin{equation}
    \label{eq:Yupdatedvs.Yoriginal}
    Y_{\mathrm{updated}} = (1\!-\!q) \cdot \frac{A_{\mathrm{MS,1}}}{\tau_{\mathrm{MS,3}}} + q\cdot Y_{\mathrm{original}}.
\end{equation}
We can immediately see from Eq.~\ref{eq:Yupdatedvs.Yoriginal}, that if
\begin{enumerate}
    \item $q=0\rightarrow A_{\mathrm{MS,1}} = A_{\mathrm{MS,3}}$, 
    \item $q=1\rightarrow Y_{\mathrm{updated}} = Y_{\mathrm{original}}$.
\end{enumerate}
Under the assumption that $A_{\mathrm{MS,1}} = A_{\mathrm{MS,3}}$ and using the Ansatz that 
\begin{equation}
    \tau_{\mathrm{MS}} = \alpha \cdot m^{y}   \qquad \qquad \mathrm{with} \quad y < -1,
\end{equation}
\begin{equation}
    \frac{\tau_{\mathrm{MS,1}}}{\tau_{\mathrm{MS,2}}} = q^{-y},
\end{equation}
we express the functions Eq.~\ref{eq:Yoriginal} and Eq.~\ref{eq:Yupdatedvs.Yoriginal} as 
\begin{align}
    Y_{\mathrm{original}} &= 0.1 \cdot \frac{\left( 1 + q^{(1-y)} \right) } {1\!+\!q}  \frac{A_{\mathrm{MS,1}}}{\tau_{\mathrm{MS,1}}} \\
    &= f_{\mathrm{original}}(q,y) \cdot \frac{A_{\mathrm{MS,1}}}{\tau_{\mathrm{MS,1}}}, \\
    Y_{\mathrm{updated}} &= \left[  \frac{1\!-\!q}{(1\!+\!q)^{-y}} +  0.1 \cdot \frac{\left( 1 + q^{(1-y)} \right) } {1\!+\!q} \right] \cdot \frac{A_{\mathrm{MS,1}}}{\tau_{\mathrm{MS,1}}} \\
    &= f_{\mathrm{updated}}(q,y) \cdot \frac{A_{\mathrm{MS,1}}}{\tau_{\mathrm{MS,1}}},
\end{align}
respectively. Now, in Fig.\ref{fig:function_family_age3.jpg} $f_{\mathrm{original}}(q,y)$ and $f_{\mathrm{updated}}(q,y)$ are plotted for three values of $y$, i.e. $y\in (-1,-2,-3)$. We can clearly see that $f_{\mathrm{updated}}(q,y)$ produces much better results as predicted for $q=0$ and $q=1$ above. We recommend this correction from Eq.~\ref{eq:age3_correction} or similar corrections to the original Eq.~\ref{eq:age3} for everyone still using the original treatment published in \citet{Hurley2002a,Hurley2005,Glebbeek2008}.

\section{Hardware settings}
\label{sec:Hardware settings}

The simulations are executed on the \textsc{Raven} high performance computing system at the Max Planck Computing and Data Facility (MPCDF)\footnote{Raven: \url{https://www.mpcdf.mpg.de/services/supercomputing/raven}}. Here, we use from one to three full GPU computing node for the simulation, which gives the optimal performance and most cost-effective use of the computing resources. Each GPU computing node is equipped 4 Nvidia A100 GPUs (4 × 40 GB HBM2 memory per node and NVLink). Furthermore, each node has one CPU hosts, which is an Intel Xeon IceLake Platinum 8360Y with 72 CPU cores. The RAM is 512 GB per node or 256 GB of RAM depending on the node you get assigned to via the slurm system. For a detailed account including benchmarking concerning the optimal performance of the \textsc{Nbody6++GPU} code on hybrid, GPU-accelerated supercomputers, please refer to the recent review by \citet{Spurzem2023} and sources therein. Even using this state-of-the-art high performance computing (HPC) infrastructure, the simulation take up to 2 months of real time to reach a star cluster age of 5~Myr. Tests and developments on machines such as juwels-booster in Germany \footnote{Juwels-booster: \url{https://apps.fz-juelich.de/jsc/hps/juwels/booster-overview.html}}, \href{https://www.lumi-supercomputer.eu/}{Lumi}
in Finland and \href{https://leonardo-supercomputer.cineca.eu/}{Leonardo} in Italy were done. We expect more data from these infrastructures soon.

\textsc{MOCCA} simulations for this work were performed on the 'chuck' computer cluster at the Nicolaus Copernicus Astronomical Center of the Polish Academy of Sciences (CAMK PAN) in Warsaw, Poland. Taking the same initial model generated by \textsc{Mcluster} for the direct \textit{N}-body simulation, we used \textsc{MOCCA} to evolve the system up to 7 Myr within approximately 10 hours on a single CPU core. The high computational efficiency of the \textsc{MOCCA} code proved extremely valuable for iterative testing and debugging, particularly in identifying and resolving the stellar radius evolution issue discussed in Appendix~\ref{sec:radius_evolution}.

\end{appendix}

\end{document}